\documentclass[aps, prx, twocolumn,longbibliography]{revtex4-1}
\usepackage[latin1]{inputenc} 
\usepackage[T1]{fontenc}

\usepackage{graphicx}
\usepackage{framed}
\usepackage{esvect}
\usepackage{amsmath,amssymb,color,bm}
\usepackage{mathrsfs}
\usepackage[dvipsnames]{xcolor}

\renewcommand{\d}{\mathrm{d}}
\renewcommand{\v}{\mathbf{v}}
\renewcommand{\r}{\mathbf{r}} 
\newcommand{\q}{\mathbf{q}}
\renewcommand{\Im}[1]{\mathrm{Im} \, [ #1]}
\newcommand{\rev}[1]{{\color{black} {#1}}}

\newcommand\tb[1]{\textbf{#1}}

\newcommand\mc[1]{\mathcal{#1}}
\newcommand\e[1]{\cdot 10^{#1}}
\newcommand\dd{\textnormal{d}}
\newcommand\beq{\begin{equation}}
\newcommand\eeq{\end{equation}}
\newcommand\beqa{\begin{eqnarray}}
\newcommand\eeqa{\end{eqnarray}}

\newcommand\tn[1]{\textnormal{#1}}
\def\Ha{\mathcal{H}}

\def\ch{\tn{cosh}}

\def\coth{\tn{cotanh}}
\def\tr{\tn{Tr}}
\def\im{\tn{Im}}

\begin{document}
\title{Quantum feedback at the solid-liquid interface: \\ flow-induced electronic current and \rev{its negative contribution to} friction}

\author{Baptiste Coquinot${}^{1,2}$}
\author{Lyd\'eric Bocquet${}^1$}
\author{Nikita Kavokine${}^{2,3}$}
\email{nkavokine@flatironinstitute.org}

\affiliation{${}^1$\!\!\! Laboratoire de Physique de l'\'Ecole Normale Sup\'erieure, ENS, Universit\'e PSL, CNRS, Sorbonne Universit\'e, Universit\'e Paris Cit\'e, 24 rue Lhomond, 75005 Paris, France\\
${}^2$\!\!\! Center for Computational Quantum Physics,
Flatiron Institute, 162 5$^{th}$ Avenue, New York, NY 10010, USA \\
${}^3$ \!\!\! Department of Molecular Spectroscopy, Max Planck Institute for Polymer Research, Ackermannweg 10, 55128 Mainz, Germany}

\date{\today}

\begin{abstract}
An electronic current driven through a conductor can induce a current in another conductor through the famous Coulomb drag effect. Similar phenomena have been reported at the interface between a moving fluid and a conductor, but their interpretation has remained elusive. Here, we develop 
a quantum-mechanical theory of the intertwined fluid and electronic flows, taking advantage of the non-equilibrium Keldysh framework. We predict that a globally neutral liquid can generate an electronic current in the solid wall along which it flows.  This hydrodynamic Coulomb drag originates from both the Coulomb interactions between the liquid's charge fluctuations and the solid's charge carriers, and the liquid-electron interaction mediated by the solid's phonons. 
We derive explicitly the Coulomb drag current in terms of the solid's electronic and phononic properties, as well as the liquid's dielectric response, a result which quantitatively agrees with recent experiments at the liquid-graphene interface. 
Furthermore, we show that the current generation counteracts momentum transfer from the liquid to the solid, leading to a reduction of the hydrodynamic friction coefficient through a quantum feedback mechanism. Our results provide a roadmap for controlling nanoscale liquid flows at the quantum level, and suggest strategies for designing materials with low hydrodynamic friction.

\end{abstract}

\maketitle 

\section{Introduction}

\begin{figure*}
\centering \includegraphics[scale=1]{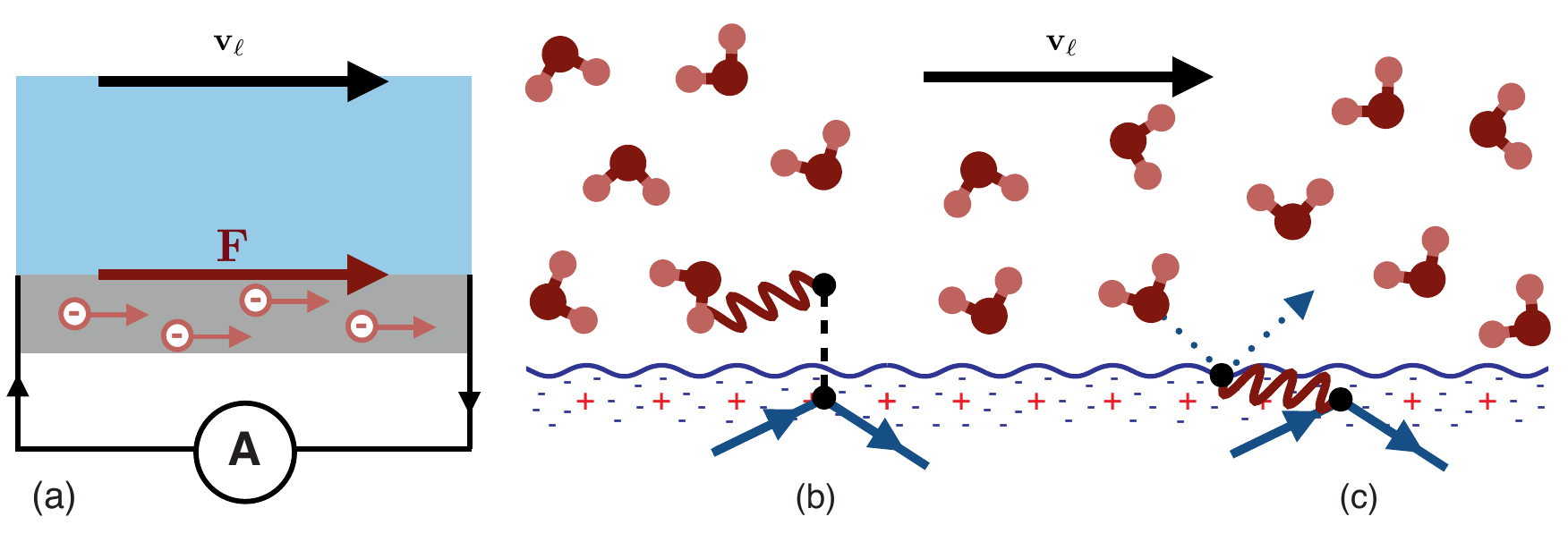}
\caption{\textbf{Mechanisms of hydrodynamic Coulomb drag}. (a) Schematic representation of the phenomenon under study: a liquid flow induces an electronic current in a solid along which it flows. (b)-(c) Schematic of the two limiting mechanisms of current generation. In (b), electrons (full lines) are driven directly by Coulomb interactions (dashed line) with liquid charge fluctuations (hydrons, wavy line). In (c) electrons are driven by phonons (wavy line), that are excited through hydrodynamic friction with the liquid. }
\label{fig1}
\end{figure*}

New functionalities in nanoscale fluid transport have been achieved by exploiting analogies with condensed matter phenomena. The analogy between surface charge in a nanochannel and doping in a semiconductor~\cite{Bocquet2010} has led to the development of nanofluidic diodes~\cite{Vlassiouk2007,Picallo2013}, and transistors~\cite{Karnik2006}; the similarity between ionic and electronic Coulomb interactions~\cite{Kavokine2021} allows for ionic Coulomb blockade~\cite{Kavokine2019,Tanaka2017,Feng2016a,Chen2021}. More recently, it has been suggested that -- beyond mere analogies --, nanofluidic transport can directly couple to electronic effects within the channel wall, {as the solid-liquid interface can host fluctuation-induced electromagnetic phenomena~\cite{Volokitin2007, Volokitin2017} : energy and momentum transfer mediated by interfacial charge fluctuations}. For instance, it has been predicted that a quantum contribution to hydrodynamic friction results from the interaction of charge fluctuations in the liquid with electronic excitations in the solid~\cite{Kavokine2022}. 

A more straightforward example of such a liquid-electron coupling is apparently provided by the numerous observations of a liquid-flow-induced current (or voltage drop) within a solid wall~ \cite{Ghosh2003, Newaz2012, HoLee2013, Yin2014, Yang2018, Rabinowitz2020, Marcotte2022}. The mechanisms proposed to explain the current generation include the buildup of a streaming potential~\cite{Cohen2003,Newaz2012}, charging/discharging of a pseudo capacitance~\cite{Yin2014,Yang2018,Park2017}, or adsorbed/desorbed ion hopping~\cite{Persson2004, Dhiman2011}. In all these cases, the liquid is in fact simply acting on the solid as an average external potential. Yet, some of the most recent experimental results~\cite{HoLee2013,Marcotte2022} cannot be explained by the above-mentioned mechanisms. \rev{In ref.~\cite{HoLee2013}, the generation of an open-circuit voltage across a millimeter-sized graphene sample due to the flow of various liquids was reported -- external potential effects could not account for these results as the liquids were ion-free. In ref.~\cite{Marcotte2022}, we have carried out analogous experiments with a thousand times smaller sample dimensions, which exclude any mesoscale charge inhomogeneities. Our observation of a liquid-flow-induced electronic current thus suggests} to examine the possibility of "intrinsic" current generation {in the framework of fluctuation-induced electromagnetic phenomena}, which would be analogous to the Coulomb drag effect in condensed matter physics (Fig. \ref{fig1}a). 

In Coulomb drag, an electric current driven through a conductor induces a current in a nearby -- yet electrically insulated -- conductor. This is due to charge fluctuations in the driven conductor creating particle-hole excitations with non-zero momentum in the passive conductor, which results in a current, provided that the conductors are not particle-hole symmetric~\cite{Narozhny2016}. A similar process could occur at the interface between a solid and a flowing liquid: excitations in the solid would then be generated through Coulomb interactions with (collective) charge fluctuations in the liquid, that we shall in the following call \emph{hydrons} (Fig. \ref{fig1}b). {Pioneering attempts at describing such a mechanism have been made by Volokitin and Persson. They applied the general theory of momentum transfer between two media through evanescent electromagnetic waves to determine the trans-resistivity of two closely-spaced solids~\cite{Volokitin2001}. Later, they expressed the electric field induced in a solid by a flowing ionic solution~\cite{Volokitin2008}, and studied the effect of substrate optical phonons on the electric current in a graphene sample~\cite{Volokitin2011}. Their approach, however, remains macroscopic, in the sense that the interacting media are described at the level of their dielectric functions. It is thus unable to reproduce the most general theoretical result that has been established for solid-solid Coulomb drag~\cite{Narozhny2016}, and a more microscopic theory is thus also required to rigorously describe the solid-liquid analogue.}

An alternative Coulomb drag mechanism, where the liquid-electron interaction is mediated by the solid's acoustic phonons (in short, \emph{phonon drag}, see Fig. \ref{fig1}c), has been first proposed by Kr\`al and Shapiro, and formalized in a Boltzmann equation framework~\cite{Kral2001}. Phonon drag has been invoked, for instance, to account for the experimental results of ref.~\cite{HoLee2013}. {It has also been suggested as a mechanism of momentum transfer between two fluids separated by a solid wall \cite{Andreev1971}.} 

In this paper, we develop a microscopic theory of electronic current generation at a solid-liquid interface, that includes both types of solid-liquid interactions. \rev{Accounting for the physics at play required us to adopt a new theoretical strategy, at odds with existing approaches to comparable problems. In particular, a mesoscopic description of the solid at the level of its dielectric function~\cite{Volokitin2001,Volokitin2008,Volokitin2011} was insufficient, since it is imprecise with regard to the mechanisms by which electrons relax their momentum. Conversely, descriptions based on the Boltzmann equation for the electrons~\cite{Kral2001,Narozhny2016} accurately capture the electron relaxation mechanisms, but fail to systematically include electron-electron interactions, thus missing the effect of the solid's plasmon modes, which can play a key role in solid-liquid systems~\cite{Yu2023}. }

\rev{To overcome these limitations, we made use of the non-equilibrium Keldysh framework of many-body quantum theory~\cite{Kavokine2022}. Our diagrammatic description allows for the inclusion of all interactions in a systematic way, possibly in the framework of numerical methods such as diagrammatic Monte Carlo~\cite{Prokofev1998,Gull2011,Bertrand2019}. Proceeding with controlled approximations, we derived an explicit expression for the electronic current generated by liquid flow. Our description of electron relaxation being fully microscopic, we could compare the relative importance of phonon-mediated and direct Coulomb solid-liquid interactions and unveil their interplay.}  Strikingly, we found that the current generation triggers a \emph{quantum feedback} mechanism at the solid-liquid interface, that reduces the total hydrodynamic friction. \rev{Our results account qualitatively for the flow-induced voltage reported in ref.~\cite{HoLee2013} and agree quantitatively with our own flow-induced current measurements~\cite{Marcotte2022}}. 

The paper is organized as follows. In Sec. II we present our model and state the main results. In Sec. III, the formal derivation is carried out. The reader interested only in the physical outcomes may skip directly to Sec. IV, where we evaluate explicitly the Coulomb drag current \rev{and compare it to the experimental results of ref.~\cite{Marcotte2022}}. In Sec. V, we derive the hydrodynamic friction renormalization resulting from the current generation; finally Sec. VI establishes our conclusions.

\paragraph*{Units and conventions.} We set the Boltzmann constant $k_{\rm B}=1$ (that is, we express the temperature in energy units), but otherwise use SI units throughout the text. Matrices are denoted with bold capital letters. We use the following convention for the $d$-dimensional Fourier transform:
\beqa 
\hat{F}(\tb{q},\omega)&=&\int \dd^d\tb{r}\dd t\, F(\tb{r},t)e^{-i\tb{q}\cdot\tb{r}+i\omega t},\nonumber\\
F(\tb{r},t)&=&\int \frac{\dd^d\tb{q}\dd\omega}{(2\pi)^{d+1}}\, \hat{F}(\tb{q},\omega)e^{i\tb{q}\cdot\tb{r}-i\omega t}.\nonumber
\eeqa

\section{Model and main results}

We consider a two-dimensional solid occupying the plane $z=0$, in contact with a semi-infinite liquid occupying the half-space $z>0$, and flowing along the $x$ direction with a velocity $\tb{v}_{\rm \ell}$, as depicted in Fig. \ref{fig1}a. The system is at temperature $T = 300~\rm K$ (or 26 meV). The flow field $\tb{v}_{\rm \ell}$ is assumed uniform in the interfacial liquid layer~\cite{Kavokine2022}. The liquid interacts with the solid through Coulomb forces. The corresponding \emph{electron-hydron} Hamiltonian is 
\beq\label{Heh}
\mc{H}_{\rm h/e}(t) = \int\dd\tb{r}\dd\tb{r}_{\rm e}\, n_{\ell}(\tb{r}-\tb{v}_{\ell} t,t)V^{\rm C}(\tb{r}-\tb{r}_{\rm e})n_{\rm e}(\tb{r}_{\rm e},t), 
\eeq
where $n_{\ell}$ and $n_{\rm e}$ are the liquid and solid charge density, respectively, and $V^{\rm C}(\r) = e^2/(4\pi \epsilon_0 r)$ is the Coulomb potential. Following~\cite{Kavokine2022}, we treat $n_{\ell}$ as a free bosonic field, whose correlation functions are related to the liquid's dielectric response. An additional liquid-solid interaction originates from short-range repulsion forces which result in "classical" hydrodynamic friction~\cite{Kavokine2022,Bocquet2007}. We will assume that this hydrodynamic friction transfers momentum to the solid's acoustic phonons~\cite{Kral2001}: those phonons then acquire a non-zero average momentum. This effect can be modeled by adding a shift $\v_{\rm ph}$ to the phonon (sound) velocity. The electron-phonon interaction Hamiltonian is then of the form 
\beq\label{Heph}
\mc{H}_{\rm ph/e}(t) = \int\dd\tb{r}\dd\tb{r}_{\rm e}\, \phi_{\rm ph}(\r- \r_{\rm e} - \v_{\rm ph} t)  n_{\rm e}(\tb{r}_{\rm e},t). 
\eeq
Here, $\phi_{\rm ph}$ is proportional to the local lattice displacement; its exact expression depends on the particular solid under consideration. The "phonon wind" velocity $\v_{\rm ph}$ is not known \emph{a priori} and it will be determined self-consistently by establishing the system's momentum balance (see Sec. \ref{wind}). 

The treatment of this model within non-equilibrium perturbation theory allows us to obtain two key analytical results. First, we obtain an explicit expression for the electronic current density $\mathbf{j}$ induced in the solid by the liquid flow for each electronic band: 
\begin{equation}\label{current}
\langle\tb{j} \rangle =  \frac{e\hbar}{T} \int_{-\infty}^{+\infty}\frac{ \d\omega}{2 \pi}\int_0^{+\infty} \frac{\d q }{2 \pi}   \, \frac{q^2 \left(\nabla_{q} \xi_q \right)\v_{\rm e}(q)}{\ch^2\left[\frac{\hbar(\omega + \xi_q)}{2T}\right]} \frac{\tau_q }{1+ \tau_q^2 \omega^2}.
\end{equation}
Here $\tau_q$ is the lifetime of the quasiparticle at energy $\hbar\xi_\q=u_\q-\mu$ where $u_\q$ is the band dispersion and $\mu$ is the chemical potential. $\v_e$ is the electron "wind velocity", which is a linear combination of the phonon wind velocity $\v_{\rm ph}$ and the hydron wind velocity (or simply, liquid flow velocity) $\v_{\ell}$: 
 \beq\label{ve0}
\v_{\rm e}(q)=\frac{\tau_\q}{\tau^{\rm ph/e}_\q}\tb{v}_{\rm ph}+\frac{\tau_\q}{\tau^{\rm h/e}_\q}\tb{v}_{\rm \ell}. 
\eeq
$1/\tau^{\rm ph/e}_\q$ and $1/\tau^{\rm h/e}_\q$ are the phonon and hydron contributions to the total quasiparticle scattering rate $1/\tau_{\q}$. This result is valid under a few reasonable assumptions on the electronic self-energy (see Sec. III), and as long as the electronic structure has no band crossings close to the Fermi level. 

Second, we predict a reduction of the hydrodynamic friction $\lambda$ coefficient due to the current generation. We recall that the solid-liquid friction force is given by $\mathbf{F} = - \lambda \mathcal{A} \v_{\ell}$, where $\cal A$ is the surface area. Accounting for the current generation, $\lambda$ is modified from its "bare" value $\lambda_0$ according to 
\beq\label{lambda}
\lambda = \frac{\lambda_0}{1+(\lambda_{\rm h/ph} + \frac{\tau}{\tau^{\rm ph/e} }\lambda_{\rm h/e}) / \lambda_{\rm um} }. 
\eeq
Here, $\lambda_{\rm h/ph}$ and $\lambda_{\rm h/e}$ are the phononic and electronic contributions to the fluctuation-induced solid-liquid friction~\cite{Kavokine2022}; $\lambda_{\rm um} = 3 \zeta(3) T^3/(2\pi \hbar^2 c^4 \tau_{\rm um})$ has the dimension of a friction coefficient, and is expressed in terms of the sound velocity $c$ in the solid, and the typical phonon lifetime $\tau_{\rm um}$.
 We demonstrate that this friction reduction is a quantum effect, that takes its roots in the solid's electronic excitations.

These results are derived in detail in the following sections. In Sec. IV, we evaluate the flow-induced current for different material systems, and successfully compare our predictions with experimental data~\cite{Marcotte2022}. In Sec. V, we show that the correction to hydrodynamic friction in Eq.~\eqref{lambda} can be non-negligible, and leads to significant hydrodynamic slippage in systems where it would not typically be expected.

\section{Non-equilibrium perturbation theory}

\subsection{Description in the Keldysh framework} 

We describe the system's dynamics in terms of three types of real-time Green's functions: the Retarded, Advanced and Keldysh Green's functions, defined, for both bosons and fermions, according to 
\beqa\label{definition_Green}
\left\{ \begin{array}{l}
G^{\rm R}(\tb{r},t,\tb{r}',t')= -i\theta(t-t')\langle[\psi(\tb{r},t),\psi^\dagger(\tb{r}',t')]_s\rangle,\\
G^{\rm A}(\tb{r},t,\tb{r}',t')= i\theta(t'-t)\langle[\psi(\tb{r},t),\psi^\dagger(\tb{r}',t')]_s\rangle,\\
G^{\rm K}(\tb{r},t,\tb{r}',t')=-i\langle[\psi(\tb{r},t),\psi^\dagger(\tb{r}',t')]_{-s}\rangle,
\end{array}\right.
\eeqa 
where $\psi^{\dagger}$ and $\psi$ are the particles' creation and annihilation operators, and $[A,B]_\pm=AB\pm BA$, $s$ being $+$ for fermions and $-$ for bosons. The Retarded and Advanced Green's functions contain information on the system's elementary excitations. For non-interacting electronic quasiparticles in a translationally-invariant system at equilibrium, the Fourier-transformed Green's functions are 
\beq
G^{\rm R,A}_0(\tb{q},\omega)=\frac{1}{\omega-\xi_\tb{q} \pm i0^+}
\label{GReq}
\eeq
 where $\hbar\xi_\tb{q}=u_\tb{q}-\mu$ is the quasiparticle energy: $u_\tb{q}$ is the band dispersion and $\mu$ is the chemical potential. The Keldysh Green's function contains information on the quasiparticle distribution. At equilibrium, it satisfies the fluctuation-dissipation theorem: 
 \beq
 G^{\rm K}(\tb{q},\omega)=\frac{2 i}{f(\omega)}\im[G^{\rm R}(\tb{q},\omega)]
\label{GKeq}
 \eeq
 where $f(\omega)=\coth\left(\frac{\hbar\omega}{2T}\right)$. Given its importance for the subsequent discussion, we recall the derivation of this result in Appendix \ref{appendix_FDT}. For non-interacting electrons, Eqs.~\eqref{GReq} and \eqref{GKeq} yield 
\beq
 G^{\rm K}(\tb{q},\omega)=(2n_{\rm F}(\omega) -1) \times  2 i \pi \delta(\omega- \xi_{\q}),
\eeq
where we recover indeed the Fermi-Dirac distribution $n_{\rm F}(\omega) = 1/(e^{\hbar\omega/T} + 1)$.

We use the letter $D$ to denote bosonic Green's functions. For free bosons (such as phonons) with dispersion $\omega_{\q}$ at equilibrium, 
\beq
D^{\rm R,A} = \frac{\omega_\q}{(\omega\pm i0^+)^2-\omega_\q^2}. 
\eeq
The bosonic fluctuation-dissipation theorem reads (see Appendix \ref{appendix_FDT})
  \beq\label{DFD}
  D^{\rm K}(\tb{q},\omega)=2 if(\omega)\im[D^{\rm R}(\tb{q},\omega)].
  \eeq
  so that, at equilibrium 
  \beq
  D^{\rm K} = (2n_{\rm B}(\omega) +1) \times   i \pi \left[\delta(\omega- \omega_\q)-\delta(\omega+ \omega_\q)\right], 
  \eeq
 where we recover the Bose-Einstein distribution $n_{\rm B}(\omega) = 1/(e^{\hbar\omega/T} - 1)$. 
 
 The Keldysh Green's functions are therefore the analogues of the occupation distribution functions in the approximate Boltzmann formalism. They will be key in determining the non-equilibrium state of the system. Indeed, as shown in Appendix \ref{appendix_current}, the current density (within one electronic band) is given by 
 \beq
 \langle \mathbf{j} \rangle = 2i e \int \frac{\d \q \d \omega}{(2\pi)^3} (\nabla_{\q} \xi_{\q})   G^{\rm K} (\q,\omega). 
 \label{current0}
 \eeq
In addition, the non-equilibrium density-density response function, which will be required for obtaining the correction to the hydrodynamic friction coefficient, can be computed starting from the non-equilibrium Green's functions (see Appendix \ref{appendix_susceptibility}). 

\begin{figure*}
\centering \includegraphics[width=0.8\textwidth]{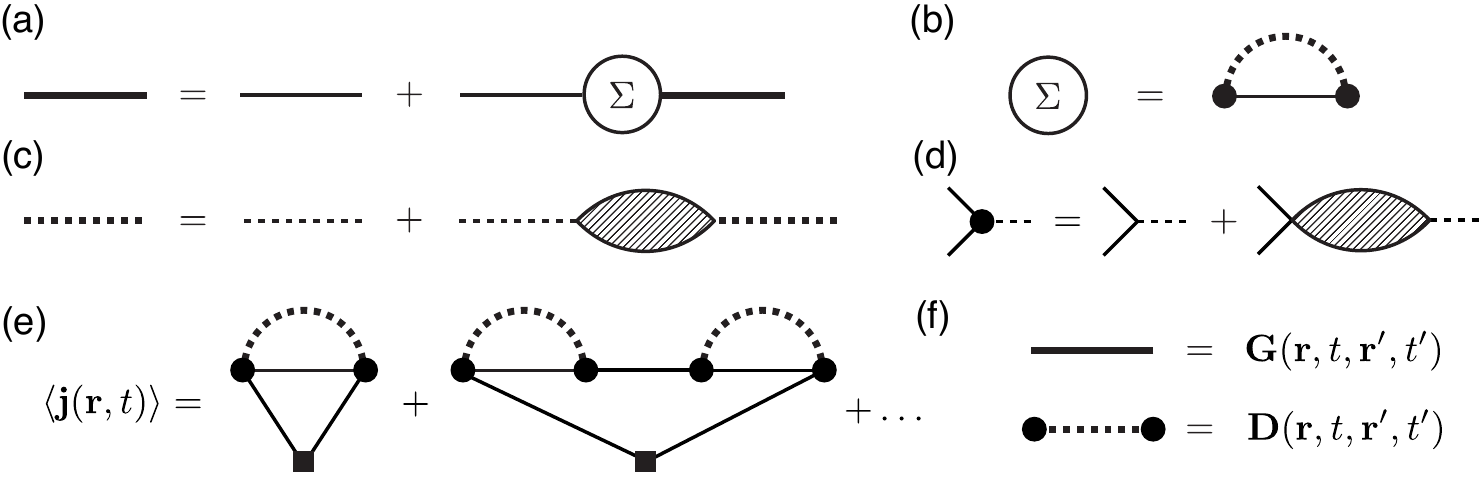}
\caption{\textbf{Non-equilibrium diagrammatic expansion.} (a) Dyson equation for the matrix Green's function $\tb{G}$ (thick line). The thin line is the bare Green's function $\tb{G}^0$.  (b) First order self-energy diagram, which is computed for each of the electron-boson interactions. (c) Dyson equation for the random phase approximation (RPA) screening of the boson propagator (dashed line). (d) Dyson equation for the boson-electron vertex, within RPA. (e)  Diagrammatic expansion for the electric current, which is related to the Keldysh Green's function at equal point in space and time. The square represents a gradient operator (multiplication by $\q$ in momentum space). (f) Notation for the electron and boson Green's functions. We have absorbed the RPA renormalisation of the vertices into the definition of the bosonic propagator. }
\label{fig2}
\end{figure*}

\subsection{Dyson equation}
Our task is now to compute the non-equilibrium Green's functions in the presence of the perturbations $\mc{H}_{\rm h/e}(t)$ and $\mc{H}_{\rm ph/e}(t)$. In the Keldysh formalism, we consider the matrix Green's function 
\beqa\label{definition_Green0}
\tb{G}=\left( \begin{array}{cc}
G^{\rm R} & G^{\rm K}\\
0 & G^{\rm A}
\end{array}\right).
\eeqa 
The perturbation series may be partially resummed by introducing a (matrix) self-energy $\bf \Sigma$. The Green's function then satisfies the non-equilibrium Dyson equation 
\beq
\bf G = G_0 + G_0 \otimes \boldsymbol{\Sigma} \otimes G, 
\label{dyson}
\eeq
which is represented diagrammatically in Fig. \ref{fig2}a. Here, $\otimes$ represents convolution in space and time, as well as matrix multiplication. We assume that the system is translationally invariant parallel to the interface, and that it has reached a steady state: we may then Fourier-transform Eq.~\eqref{dyson}. With the convolutions becoming products in Fourier space, and using that $G_0^{\rm A}(\q,\omega) = G_0^{\rm R} (\q,\omega)^*$ and $\Sigma^{\rm A}(\q,\omega) = \Sigma^{\rm R} (\q,\omega)^*$, we obtain 
\beqa\label{G2R}
G^{\rm R,A}(\q,\omega) &=& \dfrac{G_0^{\rm R,A} - |G_0^{\rm R}|^2 \Sigma^{\rm A,R}}{\left|1- G_0^{\rm R} \Sigma^{\rm R}\right|^2},\\
G^{\rm K}(\q,\omega) &=& \dfrac{G_0^{\rm K} + |G_0^{\rm R}|^2 \Sigma^{\rm K}}{\left|1- G_0^{\rm R} \Sigma^{\rm R}\right|^2}. \label{G2K}
\label{dysonGK}
\eeqa
Using Eqs. \eqref{GReq} and \eqref{GKeq}  for the equilibrium Green's functions, we find that the first term in Eq.~\eqref{dysonGK} vanishes if the self-energy is non-zero. Then, recalling Eq.~\eqref{current0}, we obtain a first very general expression for the flow-induced electric current (within a given electronic band): 
 \beq
 \langle \mathbf{j} \rangle = 2i e \int \frac{\d \q \d \omega}{(2\pi)^3} (\nabla_{\q} \xi_{\q})  \dfrac{|G_0^{\rm R}|^2 \Sigma^{\rm K}}{\left|1- G_0^{\rm R} \Sigma^{\rm R}\right|^2}. 
 \label{current2}
 \eeq
 We note that this expression is valid far from equilibrium, and that it allows for systematic inclusion of electron-electron interactions.

\subsection{Non-equilibrium self-energy}\label{SE}
In order to proceed, we need to evaluate the non-equilibrium self-energy $\boldsymbol{\Sigma}$. It contains contributions from both the electron-phonon and electron-hydron interaction. {We do not consider here any contribution of electron-electron interactions to the self-energy. We expect this to be reasonable as long as electron-phonon and electron-hydron scattering dominate electron-electron scattering, which is typically the case at room temperature~\cite{Polini2020}}. Neglecting diagrams where the phonon and hydron propagators cross, they may be computed separately: $\boldsymbol{\Sigma} = \boldsymbol{\Sigma}_{\rm h/e} + \boldsymbol{\Sigma}_{\rm ph/e}$. Furthermore, the two contributions are in fact formally identical, since both the phonons and the hydrons are free bosons coupled to the electrons. Therefore, we only need to compute a generic electron-boson self-energy, resulting from a perturbation of the form given in Eq.~\eqref{Heh}. 

We will consider a single diagram for this self-energy, as shown in Fig. \ref{fig2}b; we verify that higher order diagrams are indeed negligible under most conditions (see Appendix \ref{appendix_higherorder}). We account for the electronic screening of the bosonic propagator and boson-electron interaction vertices within the random phase approximation (Fig. \ref{fig2}c-d). We will absorb the screened vertices into the definition of the full bosonic propagator, that we denote as $\tb{D}$. The electric current, which is expressed in terms of the Keldysh Green's function evaluated at equal points in space and time, can then be represented as a sum of "ice cone" diagrams (Fig. \ref{fig2}e). These diagrams are reminiscent of the Aslamazov-Larkin diagrams~\cite{Aslamasov1968} that typically represent Coulomb drag in condensed matter systems~\cite{Kamenev1995, Narozhny2012,Kamenev2009}. Their evaluation typically involves the computation of a non-linear susceptibility (triangle diagram). In our case, this complication can be avoided, as the non-equilibrium self-energy can be readily evaluated. 

Using the Keldysh formalism Feynman rules for the boson-fermion interaction~\cite{Rammer2007} (see Appendix \ref{appendix_selfenergy}) we obtain the components of the self-energy diagram in Fig. \ref{fig2}b as:
\begin{widetext}
\beqa\label{SR}
\Sigma^{\rm R,A}(\q,\omega)  &=& - \int \frac{\d \q' \d \omega'}{(2\pi)^3} \mathcal{M}(\q-\q',\q) \left( f(\omega' -\q'\!\cdot\! \v_{\rm b}) + \frac{1}{f(\xi_{\q-\q'})} \right)\frac{ \mathrm{Im} \, [ D^R(\q',\omega'-\q'\!\cdot\!\v_{\rm b})]}{\omega-\omega'- \xi_{\q-\q'} \pm i 0^+}  , \\
\Sigma^{\rm K}(\q,\omega) &=& 2i\pi \int \frac{\d \q' }{(2\pi)^3} \mathcal{M}(\q-\q',\q)\left( 1+ \frac{f(\omega-\xi_{\q-\q'} - \q' \!\cdot\!\v_{\rm b}) }{f(\xi_{\q-\q'})} \right) \Im{D^R(\q',\omega-\xi_{\q-\q'}-\q'\!\cdot\!\v_{\rm b})},\label{SK}
\eeqa
and using $\mathrm{Im} \,[1/(\epsilon + i 0^+)] = - \pi \delta(\epsilon)$ yields
\begin{equation}\label{SR2}
\mathrm{Im} \left[\Sigma^{\rm R} (\q,\omega)\right] = \pi \int \frac{\d \q' }{(2\pi)^3} \mathcal{M}(\q-\q',\q)\left( f(\omega -\xi_{\q-\q'}-\q'\!\cdot\! \v_{\rm b}) + \frac{1}{f(\xi_{\q-\q'})} \right) \Im{D^R(\q',\omega-\xi_{\q-\q'}-\q'\!\cdot\!\v_{\rm b})} .
\end{equation}
\end{widetext}
Here $\v_{\rm b}$ is the boson wind velocity, and $\mathcal{M}(\q-\q',\q) \equiv | \langle \q-\q' | e^{-i\q' \r} | \q \rangle |^2$ are matrix elements computed between the electronic states $|\q\rangle$. The fluctuation-dissipation theorem in Eq. \eqref{GKeq} is therefore not satisfied for the non-equilibrium self-energy. Nevertheless, we may always express the Keldysh component in the form 
 \beq\label{SFD}
 \Sigma^{\rm K}(\tb{q},\omega)=\frac{2 i}{f(\omega - \q \!\cdot\!\v_{\rm eff}(\q,\omega))}\im[\Sigma^{\rm R}(\tb{q},\omega)], 
 \eeq
which defines $\v_{\rm eff}$  as a frequency and momentum dependent \emph{effective wind velocity}. 

\begin{figure*}
\centering \includegraphics[scale=1]{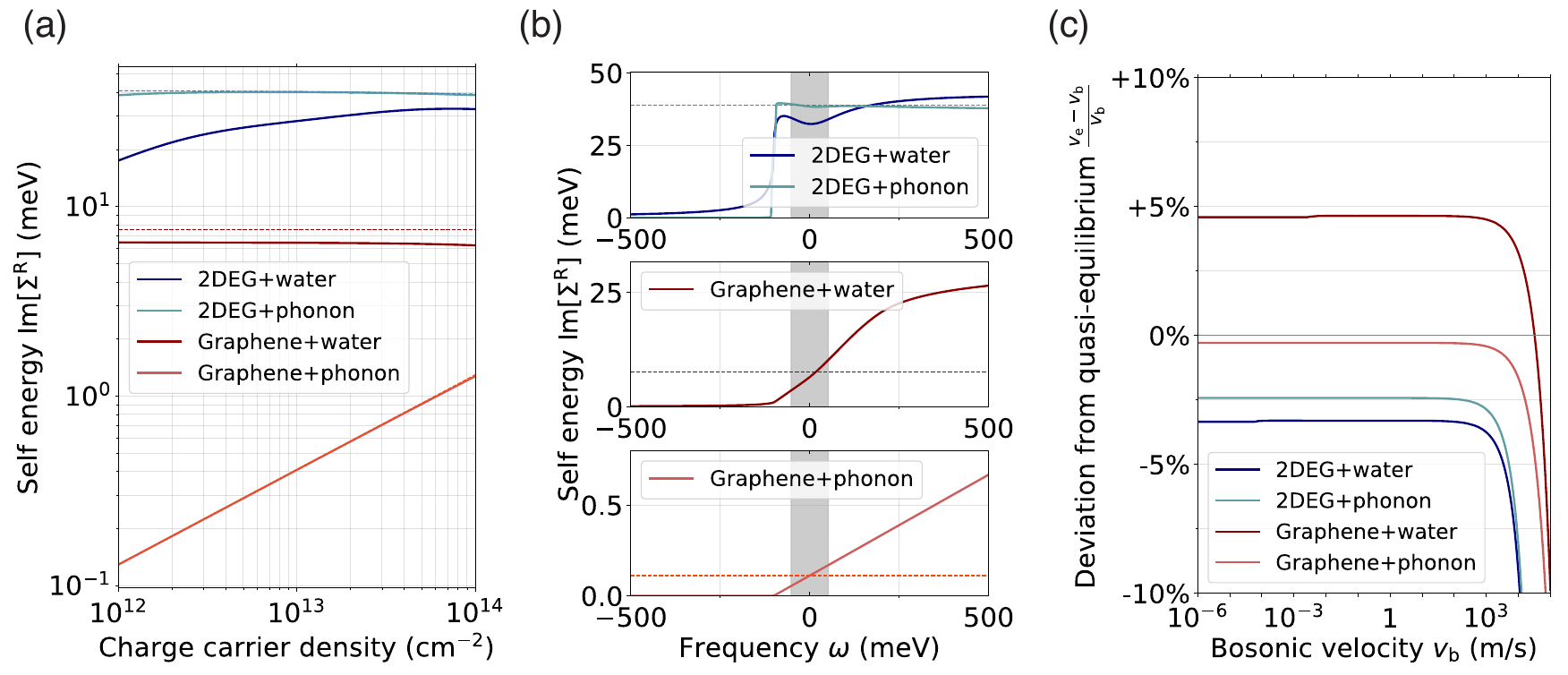}
\caption{\textbf{Non-equilibrium self energy.}  (a) Imaginary part of the electron-boson self-energy (at the Fermi level) for the four electron-boson couples discussed in the text, as a function of the charge carrier (electron or hole) density. Electron-phonon and electron-hydron scattering occur at a similar rate in the 2DEG model, whereas in graphene electron-hydron scattering is much more efficient than electron-phonon scattering. (b). Same as in (a), but plotted as a function of frequency: $\omega = 0$ corresponds to the Fermi-level. The relatively weak variations within the thermal window (grey rectangle) justify the use of the impurity approximation. (c) Deviation of the effective electronic wind velocity (defined in the text) from the bosonic wind velocity, as a function of the latter's magnitude. 
}\label{fig3}
\end{figure*}

One can now evaluate explicitly the self-energy. The calculations steps are detailed in Appendix B. In a few words, we assume for simplicity that the liquid is water, whose interfacial charge fluctuations are described in terms of a \emph{surface response function}. As demonstrated by the extensive molecular dynamics simulations of ref.~\cite{Kavokine2022}, this response function for water can be modelled as a sum of two Debye peaks, so that 
\begin{equation}\label{Dwater}
D_{\rm w}^{\rm R}(\q,\omega) =  -\frac{1}{\hbar}\frac{V_q^C}{\epsilon(q)} \sum_{k = 1,2} \frac{f_k}{1- i \epsilon(q) \omega/\omega_{\mathrm{D},k} } ,
\end{equation}
where $V_q^C=  e^2/(2\epsilon_0 q)$ is the Fourier-transformed Coulomb potential, $\omega_{\rm D, 1} \approx 1.5~\rm meV$ and $\omega_{\rm D,2} \approx 20~\rm meV$ are the Debye frequencies and $f_{1,2}$ are the corresponding oscillator strengths;
$\epsilon (q)$ is the RPA dielectric function of the electronic system, that accounts for the screening of the interaction vertices. 
{These results can be extended to any other liquid using a relevant description of their response function}. 
The acoustic phonon propagator can be written as \cite{Bruus2004}
\begin{equation}\label{Dphonon}
D_{\rm ph}^{\rm R}(\q,\omega) = \frac{1}{\hbar}V_q^{\rm ph/e} \frac{\omega_{\q}^2}{(\omega+i0^+)^2-\omega_{\q}^2}, 
\end{equation}
with $\omega_{\q} = cq$, $c$ being the phonon velocity, and $V_{\q}^{\rm ph/e}$ is the material-dependent screened electron-phonon interaction.  
For the description of the electronic system, we consider two different models: a two-dimensional electron gas (2DEG) with an effective mass $m$; and graphene, treated within the Dirac cone approximation, characterized by the constant Fermi velocity $v_{\rm F}$. We refer to Appendix B for the details on the associated electronic structure and dielectric properties. {Both models are assumed invariant by translation. The crystallographic structure of solid then appears only in the electronic propagator. More realistic models of solid may also be implemented in the theory, at the cost of higher technicality in the calculations.}

The results of these calculations are summarized in Fig. \ref{fig3}.
Fig. \ref{fig3}a shows the imaginary part of the retarded self energy at the Fermi level, $\mathrm{Im} \left[\Sigma^{\rm R}(q = k_{\rm F},\omega = 0)\right]$, computed by numerical integration according to Eq. \eqref{SR2}, where we have separated the hydron and phonon contributions. This quantity represents the scattering rate of the low-energy electronic quasiparticles. The electron-hydron interaction, as well as the electron-phonon interaction in the 2DEG are essentially screened Coulomb interactions and they yield a similar order of magnitude for the associated scattering rate. On the other hand, the electron-phonon interaction in graphene has a peculiar form (see Appendix \ref{appendix_selfenergy2}), so that the corresponding scattering rate is 2 to 3 orders of magnitudes lower. This will have an importance for the global momentum balance discussed in Sec. \ref{wind}.  

For the purpose of computing the Coulomb drag current, we need to determine how the electron-boson scattering affects the electronic distribution function, that varies typically over a scale $T$ around the Fermi level. If the boson energy is much smaller than $T$, we may approximate it as 0: the electrons then see a random static impurity potential. Within this \emph{impurity approximation}, we further assume 
\beq
\left\{
\begin{array}{ll}
&\mathrm{Im} \, [\Sigma^{\rm R} (\q,\omega)] \approx \mathrm{Im} \, [ \Sigma^{\rm R} (\q,\xi_{\q})] \equiv - 1/ \tau_{\q}, \\
&\mathrm{Re} \, [\Sigma^{\rm R} (\q,\omega)] \approx 0. 
\end{array}
\right.
\eeq
Then, as detailed in Appendix \ref{appendix_selfenergy2}, the electron-boson scattering rate can be computed as 
\begin{equation}\label{selfenergy2}
 \frac{1}{\tau_{\q}} =  \pi \frac{T}{\hbar^2} \int \frac{\d \q'}{(2\pi)^2} V_{\q-\q'} \delta(\xi_{\q} - \xi_{\q'}),  
\end{equation}
 where $V$ is the electron-boson interaction.  Fig. \ref{fig3}b shows the frequency dependence of the scattering rate (imaginary part of the self-energy) at $q = k_{\rm F}$, at a fixed chemical potential $\mu = 100~\rm meV$ for both 2DEG and graphene. 
 One observes that the variation of the self-energy in a window of width $2T$ around zero  frequency (grey rectangle in Fig. \ref{fig3}b) is relatively weak, justifying the use of the impurity approximation (dashed lines in Fig. \ref{fig3}b) in the following computations.
 
  If one neglects the angular dependence the integrand in Eq. \eqref{selfenergy2}, the scattering rate assumes an intuitive Fermi golden rule form:
 \begin{equation}\label{SEapprox}
 \frac{1}{\tau_{\q}} \approx  \pi \frac{T}{\hbar} V_{\q}N(u_\q),
\end{equation}
where $N(u_\q)$ is the density of states at energy $u_\q$. Here, the quantity $ \hbar\omega_{\rm b}V_{\q}/\mc{A}$ plays the role of the squared matrix element, and  $T/\hbar\omega_{\rm b}$ is the number of bosonic modes on which the electrons can scatter, $\hbar\omega_{\rm b}$ being the typical bosonic energy (see Appendix \ref{appendix_FGR} for a detailed derivation). 
 
 The simplified expression in Eq.~\eqref{SEapprox} allows us to understand the scalings observed in Figs.~\ref{fig3} a-b. The frequency dependence of the self-energy (Fig.~\ref{fig3}b) is roughly consistent with it being proportional to the density of states, which is independent of energy in the 2DEG ($N(u)= \theta(u)\times m/\hbar^2$) and proportional to the energy in graphene ($N(u)=\theta(u) \times 2u/(\pi\hbar)$, for the upper Dirac cone). In a 2DEG with reasonable electronic density, the screening length is much shorter than the Fermi wavelength, so that for $q\sim k_{\rm F}$, $V_{\q} \approx 1/(2 N(u_{\q})) $: we then obtain a "Planckian" scattering time  $\tau_{\q} \approx 2 \hbar / \pi T$~\cite{Zaanen2004}. This result could be expected on dimensional grounds. Indeed, once the boson energy has been neglected, the temperature is the only energy scale in the problem. The electron-boson scattering rate evaluated at the Fermi momentum is then expected to be nearly independent of electronic density, as observed in Fig.~\ref{fig3}a. The situation is different in graphene, where the screened Coulomb potential scales like $1/q$ for $q\sim k_{\rm F}$ and the density of states $N(u_\q) \propto q$ since $u_{\q} = \hbar v_{\rm F} q$. Therefore, we expect again the self-energy to weakly depend on the momentum (hence, the electronic density) for the electron-hydron scattering. For the electron-phonon scattering, where the effective potential does not depend on $q$, we expect $\tau_{\q} \propto 1/q$, so that $\tau_{k_{\rm F}} \propto n^{-1/2}$, consistently with Fig.~\ref{fig3}a.
 
 We now come back to the Keldysh component of the self-energy, which we evaluate numerically according to Eq. \eqref{SK}.  We then compute the effective electronic velocity $\v_{\rm eff}$ defined in Eq. \eqref{SFD}, at $q = k_{\rm F}$ and $\omega = 0$. The deviation of $v_{\rm eff}$ from the boson wind velocity $v_{\rm b}$ is plotted in Fig. \ref{fig3}c as a function of $v_{\rm b}$,
 for the different electron-boson couples. While the liquid flow velocity can be as low as $1~\rm \mu m/s$, we will find that phonon wind velocities can reach thousands of m/s. For this whole range of boson velocities $v_{\rm b}$ we find that $v_{\rm eff}$ remains within 5\% of $v_{\rm b}$, with a stronger deviation appearing only for $\v_{\rm b}$ in excess of $1~\rm km\cdot s^{-1}$. We may therefore safely assume $\v_{\rm eff} \approx \v_{\rm b}$, and evaluate $\Sigma^{\rm K}$ according to a \emph{quasi-equilibrium} fluctuation-dissipation theorem: 
  \beq
 \Sigma^{\rm K}(\tb{q},\omega)=\frac{2 i}{f(\omega - \q \!\cdot\!\v_{\rm b} )}\im[\Sigma^{\rm R}(\tb{q},\omega)],  
 \label{quasiFDT}
 \eeq
 which differs from the equilibrium version only by the frequency shift $\q \!\cdot\!\v_{\rm b}$. We note here the power of the Keldysh framework, which allows us to control the approximation leading to Eq.~\eqref{quasiFDT}, and potentially explore conditions where it no longer holds.

\subsection{Quasi-equilibrium state}

When Eq.~\eqref{quasiFDT} is satisfied for all electron-boson self-energies $\Sigma_j$, we will say that the system is in a \emph{quasi-equilibrium state}. Within the impurity approximation, we denote
$\im[\Sigma_j]=-1/\tau^j_{\q}$; then, $\tau_{\q}^{-1}=\sum_j(\tau^j_{\q})^{-1}$ is the total electron scattering rate at the energy $\xi_{\q}$. Eqs. \eqref{G2R} and \eqref{G2K} for the non-equilibrium Green's functions now become: 
\beqa\label{G3R}
G^{\rm R}(\q,\omega) &=& \dfrac{\sum_j\tn{Re}[\Sigma_j] +i \tau^{-1}_\q}{(\omega-\xi_{\tb{q}})^2+  \tau^{-2}_\q}\\
G^{\rm K}(\q,\omega) &=& \dfrac{2i}{f(\omega-\tb{q}\!\cdot\!\tb{v}_{\rm e}(q))}\dfrac{\tau^{-1}_\q}{(\omega-\xi_{\tb{q}})^2+  \tau^{-2}_\q} \label{G3K}
\eeqa
where the \emph{electron wind velocity} {defined as},
\beq\label{ve}
\v_{\rm e}(q)=\sum_j\frac{\tau_\q}{\tau^j_\q}\tb{v}_j,
\eeq
is a convex combination of the different boson wind velocities $\tb{v}_j$; Eq.~\eqref{G3K} is valid as long as these are small compared to the Fermi velocity. We note that we may include static impurities as an additional scatterer with zero velocity. Physically, each bosonic wind blows on the electrons through its electron-boson interaction. Each bosonic velocity contributes to the total wind velocity with a weight that is given by the corresponding electron-boson scattering rate. 
 
We therefore find that the non-equilibrium Green's function satisfies the same quasi-equilibrium fluctuation-dissipation theorem as the individual self-energies:
\beq \label{GFD}
G^{\rm K}(\tb{q},\omega)=\frac{2i}{f(\omega-\tb{q}\!\cdot\!\tb{v}_{\rm e}(q))}\im[G^{\rm R}(\tb{q},\omega)]
\eeq
Eqs.~\eqref{G3R} and \eqref{GFD} provide a complete picture of the non-equilibrium electronic state. Due to the scattering on the different bosons, the spectral function is broadened, and acquires a width $\tau_{\q}^{-1}$. In addition, the occupation of the broadened states undergoes a Doppler shift $\q\!\cdot\! \v_e$ with respect to the equilibrium Fermi-Dirac occupation. This becomes apparent if one evaluates the actual electronic density in energy-momentum space (see Appendix \ref{appendix_current}): 
\beq\label{ne}
n_{\rm e}(\tb{q},\omega)=\frac{G^{\rm K}(\q,\omega)}{2i}+\,\im[G^{\rm R}(\tb{q},\omega)],
\eeq
which reduces to 
\beq 
n_{\rm e}(\tb{q},\omega)=n_F(\omega-\tb{q}\cdot\tb{v}_{\rm e}(q))\frac{2\tau}{1+\tau^2(\omega-\xi_{\tb{q}})^2}
\eeq
where $n_F(\omega)= 1/ (e^{\hbar\omega/T}+1)$ is the Fermi-Dirac distribution. The electrons appear to acquire an average velocity equal to the electron's wind velocity, which corresponds precisely to the electric current that we evaluate in the next section. 

We make one last remark concerning the electron density-density response function. Starting from its expression in terms of the Green's functions  we demonstrate in Appendix \ref{appendix_susceptibility} that it satisfies a quasi-equilibrium fluctuation-dissipation theorem as long as the Green's functions satisfy one: 
\beq \label{chiFD}
\chi_{\rm e}^{\rm K}(\tb{q},\omega)=2if(\omega-\tb{q}\!\cdot\!\tb{v}_{\rm e}(q))\im[\chi_{\rm e}^{\rm R}(\tb{q},\omega)].
\eeq
This result will be important for the evaluation of fluctuation-induced friction forces in the next section.

\section{Flow-induced electric current}

\subsection{General expression}

Starting from the formal expression in Eq. \eqref{current0}, and using the quasi-equilibrium Green's function in Eq. \eqref{G3K}, we immediately obtain, after angular integration and to first order in the wind velocity, an explicit expression for the flow-induced electronic current: 
\begin{equation}\label{currentgeneral}
\langle\tb{j} \rangle =  \frac{e\hbar}{T} \int_{-\infty}^{+\infty}\frac{ \d\omega}{2 \pi}\int_0^{+\infty} \frac{\d q }{2 \pi}   \, \frac{q^2 \left(\nabla_{q} \xi_q \right)\v_{\rm e}(q)}{\ch^2\left[\frac{\hbar(\omega + \xi_q)}{2T}\right]} \frac{\tau_q }{1+ \tau_q^2 \omega^2}.
\end{equation}
The wind velocity $\v_e(q)$ is given by Eq. \eqref{ve}. In the presence of electron-hydron, electron-phonon and electron-impurity scattering, it explicitly writes 
 \beq\label{ve2}
\v_{\rm e}(q)=\frac{\tau_\q}{\tau^{\rm ph/e}_\q}\tb{v}_{\rm ph}+\frac{\tau_\q}{\tau^{\rm h/e}_\q}\tb{v}_{\rm \ell}
\eeq
where $1/\tau_\q=1/\tau^{\rm ph/e}_\q+1/\tau^{\rm h/e}_\q+1/\tau^{\rm im}_\q$, the last term being the impurity scattering rate. 

We thus obtain our first main result, as anticipated in Sec. II: we predict current generation in the solid due to the flow of a neutral liquid.

{ \subsection{Comparison with literature results}}

{To our knowledge, our result in Eq.~\eqref{currentgeneral} is not found in the literature. Its closest analogue is the general expression for solid-solid Coulomb drag, Eq. (15) in ref.~\cite{Narozhny2016}, since it is derived at the same level of theory. However, the solid-solid result is not directly applicable to the solid-liquid case, since it involves a non-linear current-voltage response function, which is not defined for a liquid that is assumed insulating within our model. The result in Eq. (36) is in fact simpler, because, as compared to the solid-solid case, one of the interacting fermionic systems is replaced with a free bosonic field. We note that, recently, a theory of Coulomb drag has been developed for a system of two graphene sheets where the electrons are in the hydrodynamic regime~\cite{Levchenko2022}. However, the analogy with our system is mostly semantic, as the fluctuations of an electron liquid are very different from hydron modes, which are similar to strongly damped optical phonons.}

{In the framework of fluctuation-induced electromagnetic phenomena, Volokitin and Persson have proposed an expression for the electric field $E$ induced in a 2D electron gas (2DEG) by a liquid flowing along its surface~\cite{Volokitin2008}. They considered only the contribution of direct Coulomb interactions, and did not include any phonon effects. Under the assumption that this electric field equilibrates the solid-liquid quantum friction force, they obtain $n e E = \lambda_{\rm h/ e } v_{\rm \ell}$, where $n$ is the electron density in the 2DEG and $\lambda_{\rm h/e}$ is the quantum friction coefficient (see Table 1). In order to convert this to a current density, one has to assume a Drude-like conductivity for the 2DEG, $\sigma = n e^2 \tau / m$, where $\tau$ is a momentum-independent relaxation time and $m$ is the effective mass. Then, using Ohm's law $j = \sigma E$, one obtains 
\begin{equation}
 j = \frac{e\tau}{m} \lambda_{\rm e/h} v_{\rm \ell}.
 \label{current_volokitin}
 \end{equation} 
}

{Even under the assumption of momentum-independent relaxation times and a parabolic band structure, we find that the direct Coulomb contribution in Eq.~\eqref{currentgeneral} does not reduce to Eq.~\eqref{current_volokitin}. Indeed, if we further assume a quasiparticle scattering rate that is small compared to the thermal energy $\hbar/\tau \ll T$, the Coulomb contribution in Eq. (36) becomes 
\begin{equation}
\langle j \rangle = \frac{e \tau}{m} \times \int \frac{d^2 q}{(2 \pi)^2} \hbar q   \left[ \frac{\pi q}{\ch^2\left[\frac{\hbar\xi_q}{2T}\right]} \right] \frac{1}{\tau^{\rm e/h}} v_{\ell} \equiv \frac{e \tau}{m} \lambda_{\rm e/h}^* v_{\ell}. 
\label{emission}
\end{equation} 
Comparing to Eq.~\eqref{current_volokitin}, we may identify an effective friction coefficient $\lambda_{\rm e/h}^*$, whose expression differs from the usual $\lambda_{\rm e/h}$ as obtained, for example, in~\cite{Kavokine2022} (see also Eq.~\eqref{lambda_boson}). In particular, the photon tunneling rate is given by the liquid's contribution to the electronic self-energy, $1/\tau_{\rm e/h} \equiv \mathrm{Im} \, \Sigma_{\rm h}$, rather than by the overlap of surface excitation spectra; the two expressions would in fact be equivalent only if the electrons were non-interacting. 
  }
  
{We conclude that the microscopic Keldysh-formalism approach was instrumental in obtaining a rigorous description of electronic current generation by liquid flow. We would like to stress the generality of the approach, since it formally allows for any interactions to be taken into account to any desired level of precision. Our most general result is in fact given by Eq.~\eqref{current2}, where the electric current is expressed in terms of the electronic self-energy. This self-energy may be computed within various numerical schemes (in particular, diagrammatic Monte Carlo~\cite{Prokofev1998,Gull2011,Bertrand2019}), and thus the solid-liquid Coulomb drag maybe studied within regimes where our impurity or quasi-equilibrium approximations no longer hold. 
}

\subsection{Wind velocity : global momentum balance} \label{wind}

\begin{figure*}
\centering \includegraphics[scale=1]{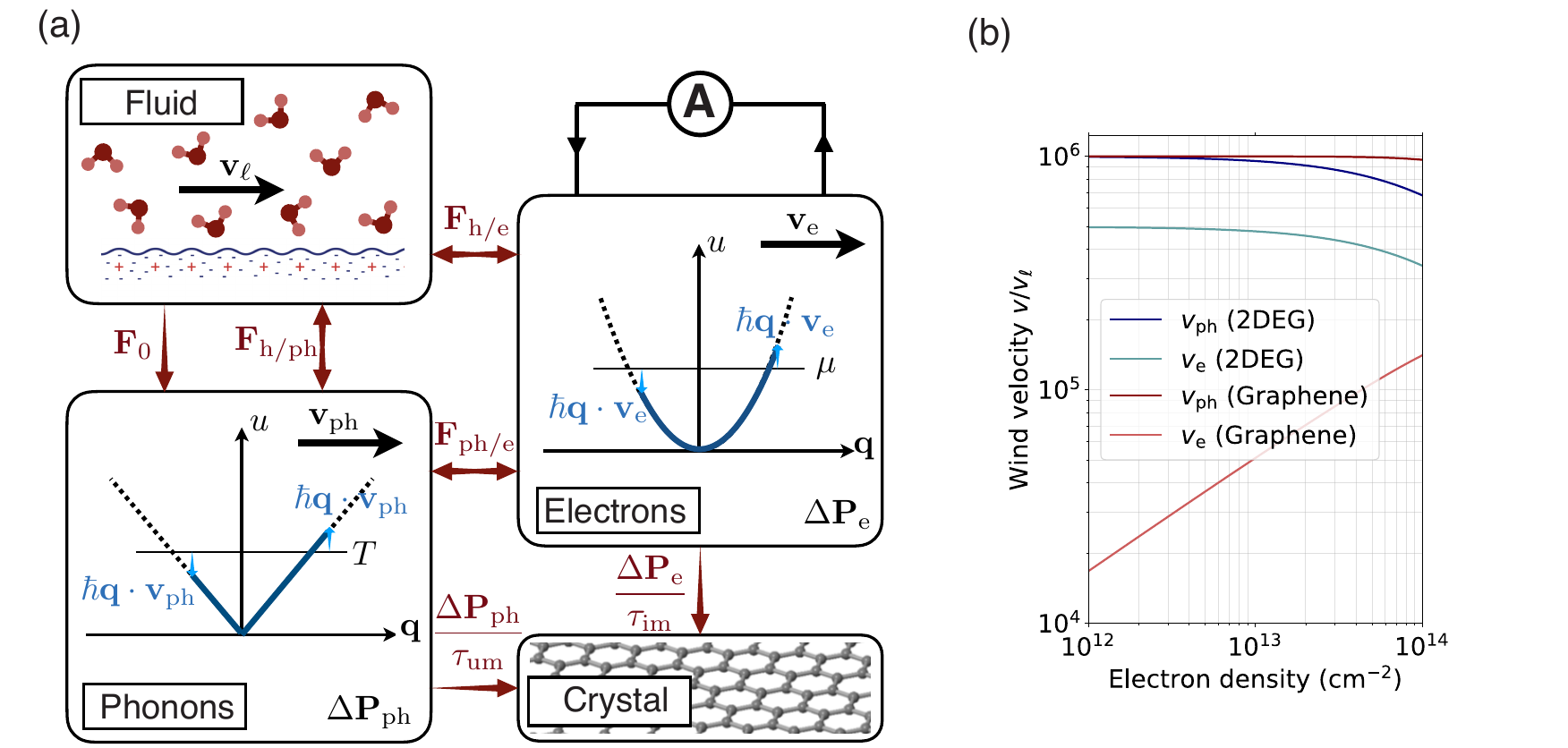}
\caption{\textbf{Momentum balance.} (a) Diagram representing the momentum fluxes in the solid-liquid system, separated into four subsystems, represented by the rectangular boxes. The liquid (flowing at velocity $\v_{\ell}$) is a momentum source, and the "crystal" is a momentum sink. The momentum fluxes in and out of the phonon and electron subsystems need to be balanced in the steady state. The phonons (electrons) accumulate a momentum $\Delta \tb{P}_{\rm ph}$ ($\Delta \tb{P}_{\rm e}$), corresponding to a Doppler shift $\hbar q v_{\rm ph}$ ($\hbar q v_e$) of their momentum distribution. $\tau_{\rm um}$ is the phonon umklapp scattering time and $\tau_{\rm im}$ is the electron impurity scattering time. 
(b) Phonon and electron wind velocities (normalized by the liquid velocity) as a function of the electronic density, for $\lambda_{\rm um} \approx 2~\rm N \cdot s \cdot m^{-3}$ and $\lambda_0 = 2 \times 10^6 ~\rm N \cdot s \cdot m^{-3}$.}\label{fig4}
\end{figure*}

\begin{table*}
\begin{tabular}{ l rcl rcl c }
    \hline \hline
    
    Interaction & \multicolumn{3}{l}{\quad\quad  Momentum transfer}  &   \multicolumn{3}{l}{\quad\quad  Force} &Scattering time \\
    \hline
    Classical friction & \quad Liquid &$\rightarrow$& Phonons \quad & $\tb{F}_0  $&=&$\lambda_0\mc{A}\tb{v}_{\ell}$ &   \\
    Phonon-hydron interaction &\quad Liquid &$\leftrightarrow$ &Phonons \quad & $\tb{F}_{\rm h/ph} $&=&$\lambda_{\rm h/ph}\mc{A}(\tb{v}_{\ell}-\tb{v}_{\rm ph})$ &   \\
           Electron-hydron interaction &\quad Electrons &$\leftrightarrow$& Liquid \quad & $\tb{F}_{\rm h/e} $&=&$\lambda_{\rm h/e}\mc{A}(\tb{v}_{\ell}-\tb{v}_{\rm e})$ & $\tau_{\rm h/e}$  \\
    Electron-phonon interaction &\quad Phonons &$\leftrightarrow$& Electrons \quad & $\tb{F}_{\rm ph/e} $&=&$\lambda_{\rm ph/e}\mc{A}(\tb{v}_{\rm e}-\tb{v}_{\rm ph})$ & $\tau_{\rm ph/e}$  \\
        Umklapp processes &\quad  Phonons &$\rightarrow$& Crystal \quad & $\tb{F}_{\rm um} $&=&$\Delta\tb{P}_{\rm ph}/\tau_{\rm um}=\lambda_{\rm um}\mc{A}\tb{v}_{\rm ph}$ & $\tau_{\rm um}$  \\ 
    Impurities &\quad  Electrons &$\rightarrow$& Crystal \quad & $\tb{F}_{\rm im} $&=&$\Delta\tb{P}_{\rm e}/\tau_{\rm im}$ & $\tau_{\rm im}$  \\ \hline 

    Total friction & \quad Liquid &$\rightarrow$& Phonons+Electrons \quad & $\tb{F} $&=&$\lambda\mc{A}\tb{v}_{\ell}=(\lambda_0+\delta\lambda)\mc{A}\tb{v}_{\ell}$ & \\ \hline \hline
\end{tabular}
\caption{\textbf{Forces in the solid-liquid system.} List of interactions that may transfer momentum between the different components of the solid-liquid system, and notations for the associated forces, friction coefficients and scattering times.}\label{able}
\end{table*} 

In order to evaluate the current in Eq.~\eqref{currentgeneral}, we require one last ingredient, which is the velocity $\v_{\rm ph}$ of the phonon wind. As mentioned in Sec. II, we evaluate it self-consistently, by enforcing momentum conservation in the solid-liquid system. 

In the stationary state, the phonons accumulate a momentum that we denote $\Delta \tb{P}_{\rm ph}$. We model this momentum accumulation by giving all the phonons an average velocity $\v_{\rm ph}$: this means shifting the phonon distribution according to 
\beq 
n_{\rm ph}(\tb{q},\omega)=n_{\rm B}(\omega-\tb{q}\cdot\tb{v}_{\rm ph})\times2\pi\delta(\omega-cq)
\eeq
where $n_{\rm B}(\omega)= 1 / (e^{\hbar\omega/T}-1)$ is the Bose-Einstein distribution and $c$ is the sound velocity. 
Then, 
\beq
\frac{\Delta \tb{P}_{\rm ph}}{\mc{A} }= \int\frac{\dd^2\tb{q}}{(2\pi)^2}\hbar\tb{q}\,n_{\rm B}\left( qc-\tb{q}\cdot\tb{v}_{\rm ph}\right), 
\eeq
which becomes, to first order in $\v_{\rm ph}$,
\beq
\frac{\Delta \tb{P}_{\rm ph}}{\mc{A} }= \frac{3\zeta(3)}{2\pi}\frac{T^3}{ \hbar^2c^4}\tb{v}_{\rm ph},
\eeq
$\zeta$ being the Riemann function. The contributions to the momentum flux in and out of the phonon system are summarized in Fig. \ref{fig4}a, and the associated notations are explicited in Table 1. The phonons receive momentum from the flowing liquid through the classical, roughness-induced contribution to the hydrodynamic friction force, $\tb{F}_0 = \lambda_0 \mc{A} \v_{\ell}$, and through the phononic contribution  $\tb{F}_{\rm h/ph}$ to fluctuation-induced friction~\cite{Kavokine2022}. In Appendix \ref{sec_force}, we extend the framework of ref.~\cite{Kavokine2022} to account for the non-equilibrium state of the solid, and show that as long as the quasi-equilibrium fluctuation-dissipation theorem (Eq. \eqref{chiFD}) holds for the system's density response functions, the fluctuation-induced friction force is proportional to the differential velocity: 
\begin{equation}
\tb{F}_{\rm h/ph} = \lambda_{\rm h/ph}\mc{A} (\v_{\ell} - \v_{\rm ph}). 
\end{equation}
We note that, conversely, $\tb{F}_0$ {-- the classical, roughness induced friction --} does not depend on the phonon velocity: indeed, it originates largely in defects on the solid's surface, which do not move even if the phonons accumulate momentum. 

The phonons lose momentum mainly through umklapp processes; we denote $\tau_{\rm um}$ the corresponding relaxation time. Formally, in an umklapp process, the interference of multiple phonons converts their momentum into a global translation of the crystal lattice.  In practice, however, the 2D material layer remains fixed, and the momentum is transferred to the underlying substrate, which we do not describe explicitly. The momentum lost by the phonons per unit time and unit area through umklapp processes is therefore 
\begin{equation}
\frac{\tb{F}_{\rm um}}{\mc{A} } = \frac{\Delta \tb{P}_{\rm ph}}{\tau_{\rm um}\mc{A} } =  \frac{3\zeta(3)}{2\pi}\frac{T^3}{ \hbar^2c^4\tau_{\rm um}}\tb{v}_{\rm ph} \equiv \lambda_{\rm um} \v_{\rm ph},
\end{equation}
where we have defined the umklapp friction coefficient $\lambda_{\rm um}$. In graphene, $\tau_{\rm um}\approx 10 ~\rm ps$ \cite{Klemens1994} and $c\approx 2\e{4}$ m/s \cite{Nika2009,Ochoa2011,Cong2019}, so that $\lambda_{\rm um}\approx 2 ~\rm N \cdot s \cdot m^{-3}$. In addition, the phonons loose momentum through quantum friction with the conduction electrons, which is analogous to the water-electron quantum friction studied in ref.~\cite{Kavokine2022}, with the Coulomb interaction being replaced by the electron-phonon interaction. This type of friction is also known as a \emph{current-induced force} in the context of nanoscale electron transport~\cite{DiVentra2008}. The corresponding momentum flux is $\tb{F}_{\rm ph/e} = \lambda_{\rm ph/e}\mc{A}  (\v_{\rm ph} - \v_{\rm e})$. In a steady state, the incoming and outgoing momentum fluxes (or forces) must compensate: 
\beq\label{balance1}
\tb{F}_{0}+\tb{F}_{\rm h/ph}= \tb{F}_{\rm um}+\tb{F}_{\rm ph/e}.
\eeq
Using Eq. \eqref{ve2} for the electronic velocity $\v_{\rm e}$, we then obtain explicitly the phonon wind velocity as 
\beq \label{vph}
\tb{v}_{\rm ph}=\frac{\lambda_0+\lambda_{\rm h/ph}+\frac{\tau}{\tau_{\rm h/e}}\lambda_{\rm ph/e}}{\lambda_{\rm um}+\lambda_{\rm h/ph}+\left(1-\frac{\tau}{\tau_{\rm ph/e}}\right)\lambda_{\rm ph/e}}\v_{\rm \ell}.
\eeq  
This formula is consistent with the roughness-induced friction (with coefficient $\lambda_0$) being a momentum source, and the umklapp processes (with equivalent friction coefficient $\lambda_{\rm um}$) being a momentum sink. We do not have a practical way of evaluating the acoustic phonon contribution $\lambda_{\rm h/ph}$ to the fluctuation-induced friction. Nevertheless, guided by the qualitative ideas of ref.~\cite{Kavokine2022}, we expect it to be very small, since it is associated with in-plane lattice displacements at wavelengths much larger than the atomic spacing, that have short-range contact interactions with the liquid. Henceforth, we will assume it to be negligible compared to $\lambda_{\rm um}$ and $\lambda_0$. The electron-phonon friction coefficient, on the other hand, is evaluated explicitly in Sec. V, and we  find that $\lambda_{\rm ph/e} \ll \lambda_{0}$ under all practical conditions; it is, however, comparable to $\lambda_{\rm um}$ at large electronic density. 
Therefore, in the absence of impurities, the phonon wind velocity may be simplified, after some rearrangements, to 
\beq \label{vph2}
\v_{\rm ph} \approx \frac{\lambda_0}{\lambda_{\rm um}+\lambda_{\rm ph/e}\tau/\tau_{\rm h/e}} \v_{\ell}.
\eeq 
Typically, on a molecularly rough surface, the roughness-induced friction coefficient is of order $\lambda_0 \sim 10^5 - 10^6~\rm N \cdot s \cdot m^{-3}$~\cite{Bocquet2010}, so that $\v_{\rm ph} \gg \v_{\ell}$: the phonon velocity is orders of magnitude larger than the flow velocity. Fig. \ref{fig4}b shows the phonon and electron wind velocities (normalized by $\v_{\ell}$) as a function of the electronic density. Under all practical conditions the phonon wind is faster than the electron wind, whose velocity is reduced by the electron-hydron interaction. This results in the electron-hydron interaction actually making a \emph{negative} contribution to the Coulomb drag, as explained in the next section.

One may draw an analogy between current generation by hydrodynamic Coulomb drag and current generation through the photoelectric effect. The former benefits from strong electron-phonon interactions, as those help transfer momentum from the liquid to the electrons. Conversely, the latter is suppressed by electron-phonon scattering, as phonon drag slows down the photo-generated charge carriers: reducing phonon drag is a key challenge in the engineering of Perovskite materials, which are the state-of-the-art for photoelectric panels \cite{Yang2015,Kentsch2018,Zhang2021}. Such a complementarity calls for further studies of the potential interplay between these two phenomena.

\subsection{Quantitative estimates: Coulomb drag vs. phonon drag}

\begin{figure*}
\centering \includegraphics[scale=1]{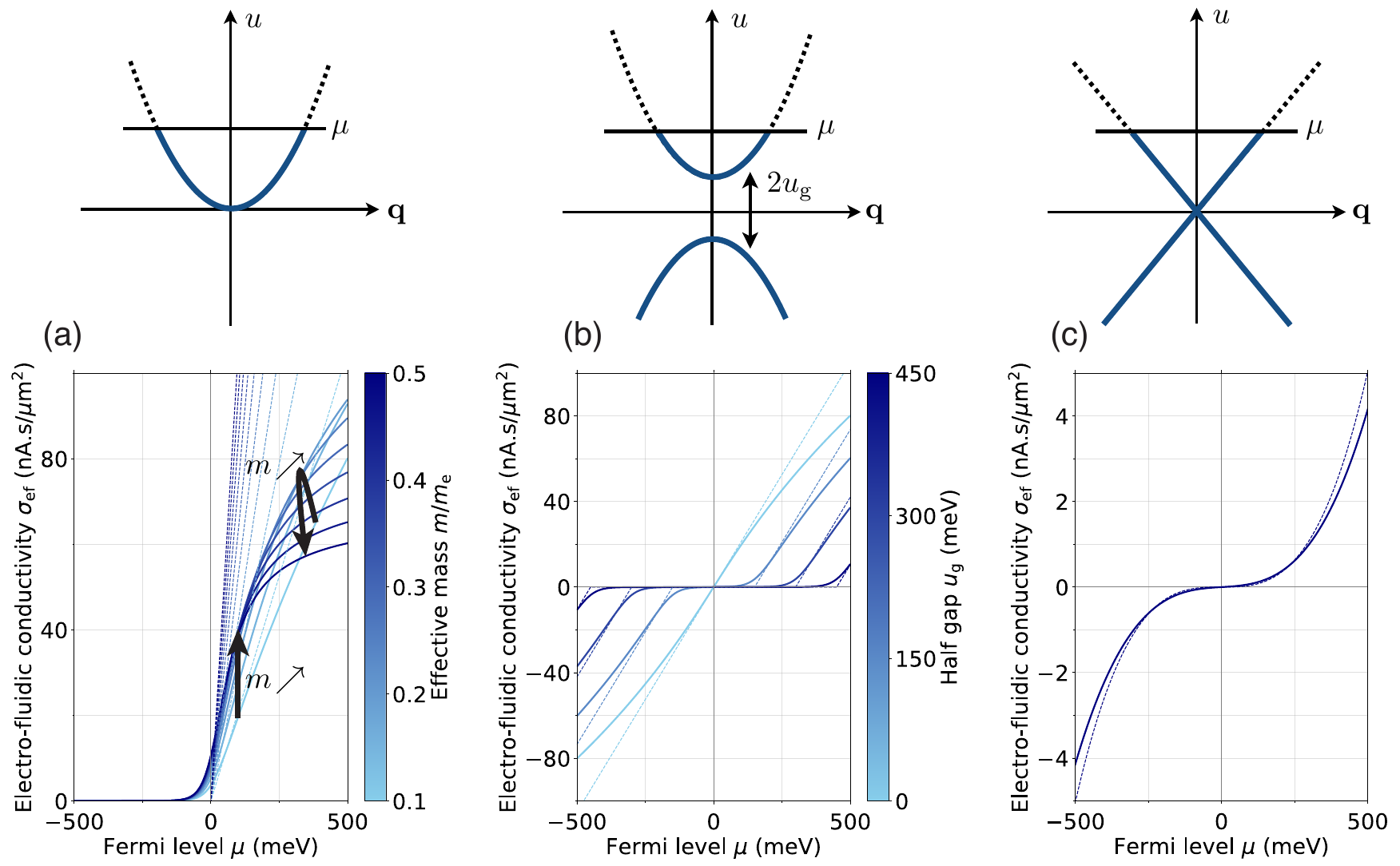}
\caption{\tb{Coulomb drag current}. Electro-fluidic conductivity $\sigma_{\rm el}=\langle\tb{j}\rangle/\v_{\rm \ell}$ as computed with Eq.~\eqref{currentgeneral}, for different models of the solid's band structure. (a) Two-dimensional electron gas, with different values of the effective mass $m$ (expressed in units of electron mass $m_{\rm e}$). (b) Semi-conductor with electron and hole masses $m=0.1m_{\rm e}$, and band gap $2u_{\rm g}$. (c) Graphene, with constant Fermi velocity in the Dirac cone approximation. In (a)-(c) we assumed $\frac{\lambda_0}{\lambda_{\rm um}}= 10^6$. The dashed lines correspond to the prediction of Eq.~\eqref{currentapprox}, with no chemical potential dependence of the electronic wind velocity.}\label{fig5}
\end{figure*}

We are now in position to evaluate the electronic current according to eqs. \eqref{currentgeneral} and \eqref{ve2}. In order to make quantitatives estimates, we will use $\lambda_{\rm um} \approx 2~\rm N \cdot s \cdot m^{-3}$ as computed for graphene in the previous section, and $\lambda_0 = 2 \times 10^6 ~\rm N \cdot s \cdot m^{-3}$, which is reasonable for water on a large area exhibiting some ripples and defects. Furthermore, we make use of the values for $\lambda_{\rm ph/e}$ computed in Sec. V B. 
Then, at not too large electronic density, $\v_{\rm ph} \approx (\lambda_0/\lambda_{\rm um}) \v_{\ell} \sim 10^6 \v_{\ell}$. 
As in Sec. III, we will consider two models for the electronic structure and electron-phonon interaction: a two-dimensional electron gas (2DEG) with an effective mass $m$, and graphene, treated within the Dirac cone approximation, characterized by the constant Fermi velocity $v_{\rm F}$. In addition we will consider a model of a direct band gap semiconductor as the combination of an electron gas and a hole gas. For the two-band systems we will evaluate the total current as the sum of the currents in the two bands, which amounts to neglecting interband scattering. Fig. \ref{fig5} shows schematics of the three band structures under consideration
 
 We define the \emph{electro-fluidic conductivity} as
 \begin{equation}
 \sigma_{\rm ef} \equiv {\langle j \rangle \over  v_{\ell}}.
 \end{equation}
We plot in Fig. \ref{fig5}a-c the electro-fluidic conductivity as a function of the chemical potential $\mu$. One first notes the difference in scaling of $\sigma_{\rm ef}$ with $\mu$ between the three model systems. These scalings are most conveniently understood in the limit of weak interactions and low temperature $\hbar/\tau_{k_{\rm F}} \ll T \ll \mu$. The first inequality means that the broadening of the electronic distribution due to electron-boson scattering is negligible compared to the thermal broadening. The second inequality means that the Fermi-Dirac distribution is well approximated by a step function. Eq.~\eqref{currentgeneral} accordingly simplifies to 
 \beq \label{currentapprox}
\langle\tb{j}\rangle\approx 2e  v_{\rm F}  N(\mu) \times \hbar k_{\rm F} \v_{\rm e}(k_{\rm F}),
\eeq
where $N(\mu)$ is the density of states at the Fermi level. The effect of the temperature appears to cancel out, leaving us with a transparent expression that is intuitive in a zero-temperature picture of Coulomb drag. The current is the electronic charge times the electronic velocity (which is the Fermi velocity), times the charge carrier density contributing to the current. The latter is the density of states at the Fermi level, times the energy range around the Fermi level in which the charge carriers can contribute to the current: this is given by the "Doppler shift" $\hbar k_F v_e$. 

In the 2DEG, the density of states $N(\mu) = m/\hbar^2$ is independent of chemical potential, and $v_{\rm F} k_{\rm F} \propto \mu$: one expects a linear scaling of $\sigma_{\rm ef}$ with $\mu$ when $\mu \gg T$. A correction to this scaling comes from the chemical potential dependence of the electron-phonon friction coefficient: $\lambda_{\rm ph/e}\propto \mu$ (see Appendix \ref{sec_lambda}), which contributes to reducing the phonon wind velocity at high $\mu$ (see Fig. \ref{fig4}b). As a consequence, the current is expected to saturate at high chemical potential. Our model for the semi-conductor is a combination of two-dimensional electron and hole gases, therefore a linear scaling of $\sigma_{\rm ef}$ is obtained for both positive and negative chemical potential (Fig \ref{fig5}b). The electro-fluidic conductivity is suppressed when the chemical potential is within the band gap because of the lack of charge carriers, similarly to the 2DEG at negative chemical potential. 
In graphene, $N(\mu) \propto \mu$, $k_{\rm F} \propto \mu$ and $v_{\rm F}$ is independent of $\mu$: we thus expect $\sigma_{\rm ef} \propto \mu |\mu|$ (the sign of $\sigma_{\rm ef}$ reflects the nature -- electron or hole -- of the charge carriers). Fig. \ref{fig5}c shows a slight deviation from this quadratic scaling, which is due to the dependence on chemical potential of the electron-phonon scattering time that contributes to the wind velocity $\v_e$ (see Fig. \ref{fig3}c and Fig. \ref{fig4}b). {We note that we consider here a simplified model of graphene that neglects interband scattering, or any effect of charge inhomogeneities that could cause a non-vanishing Coulomb drag current at charge neutrality~\cite{Gorbachev2012,Song2012}. }
 
In the electron gas model, $\sigma_{\rm ef}$ is tunable by the charge carrier effective mass. As the effective mass increases, the Fermi velocity is reduced, but the density of states and the Fermi momentum increase: the latter dominate, and overall, at low enough chemical potential ($\mu \lesssim 0.2~\rm eV$) $\sigma_{\rm ef} \propto m$. However, at larger chemical potential, the electron-phonon friction coefficient, that scales as $m^2$ (see Appendix~\ref{sec_lambda}), reduces the phonon wind velocity (see Fig. \ref{fig4}b), so that $\sigma_{\rm ef} \propto 1/m$.  These scalings, illustrated in Fig. \ref{fig5}b, suggests that flat-band materials are likely to exhibit a significant hydrodynamic Coulomb drag effect. 

At similar chemical potential, the electro-fluidic conductivity is found to be about two orders of magnitude larger in the 2DEG than in graphene. For instance at $\mu\sim 100$ meV, $\sigma_{\rm ef} \approx 20~\rm nA \cdot s \cdot \mu m^{-2}$ for the 2DEG and $\sigma_{\rm ef} \approx 0.1~\rm nA \cdot s \cdot \mu m^{-2}$ for graphene. This difference is mainly due to the wind velocity $\v_e$, which is determined according to Eq. \eqref{ve2} from the phonon wind velocity $\v_{\rm ph}$ and the flow velocity $\v_{\ell}$, with $\v_{\rm ph} \gg \v_{\ell}$ (see Fig. \ref{fig4}b). In the 2DEG the electron-phonon and electron-hydron scattering times are similar, so that $\v_e \approx \v_{\rm ph}/ 2$. In graphene, the electron-hydron scattering is much faster than the electron-phonon scattering, so that $\v_e \sim 10^{-2} \v_{\rm ph} \gg \v_{\ell}$. Thus, despite the different orders of magnitude, the phonon drag is the main driving force for the electronic current in both model systems. However, this does not imply that the electron-hydron interactions are negligible: in fact, they reduce the "bare" phonon drag current by a factor of 2 in the 2DEG and by a factor of $10^2$ in graphene, by providing a supplementary momentum relaxation pathway for the electrons. In other words, the electron-hydron interaction makes a negative contribution to the Coulomb drag. This effect is at the root of the quantum feedback phenomenon discussed in Sec. V. 

\subsection{Comparison with experiment}

\begin{figure}
\centering \includegraphics[scale=1]{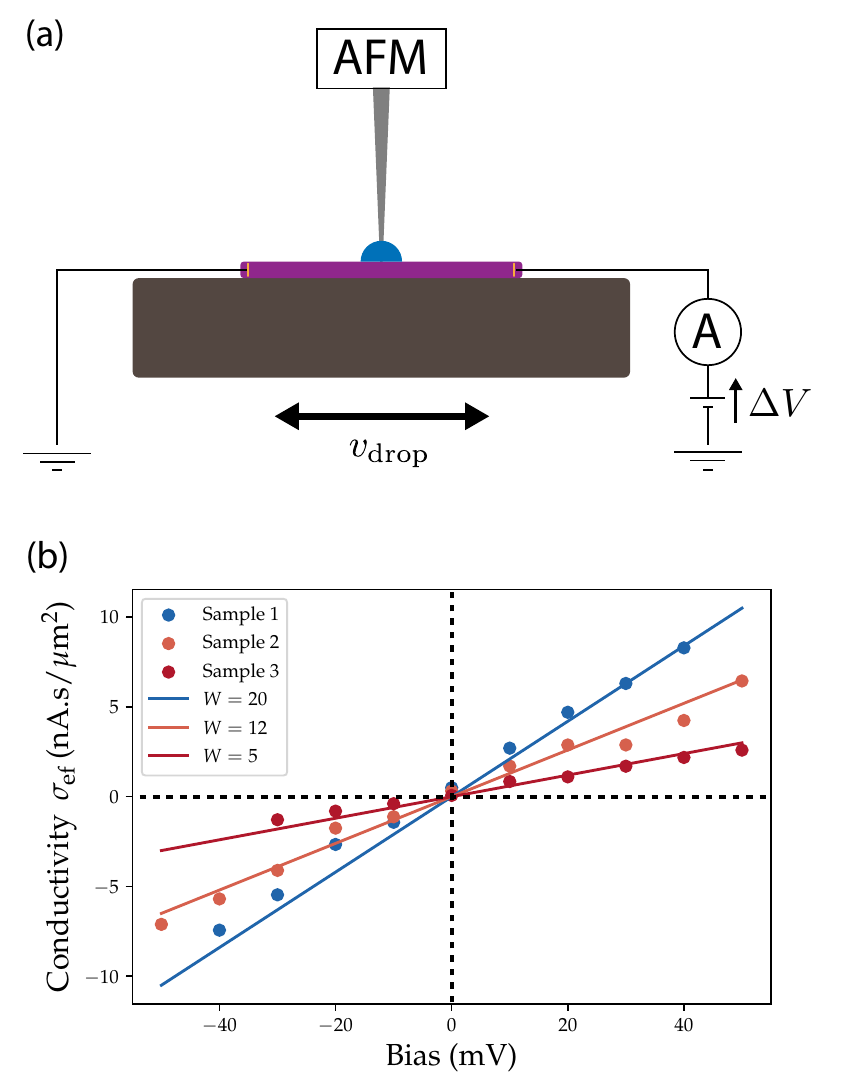}
\caption{{\textbf{Comparaison with the experiment of Marcotte \textit{et al.} \cite{Marcotte2022}}. (a) Schematic of the experimental setup. Thanks to an atomic force microscope (AFM), a liquid droplet is deposited onto the surface of a multilayer graphene sample connected to two metallic electrodes. Actuation of a piezoelectric scanner results in motion of the drop on the carbon surface at a velocity $v_{\rm drop}$. (b) Quantitative comparison between the experimental data for the electro-fluidic conductivity $\sigma_{\rm ef}$ from three different devices and theoretical predictions, parametrized by the wrinkling number $W$ that account for the wrinkle density on the sample surface.}}
\label{figexp}
\end{figure}

{Lastly, we compare the quantitative estimates obtained from our model with the results of the companion experimental paper (ref.~\cite{Marcotte2022}). In ref.~\cite{Marcotte2022}, an atomic force microscope is used to deposit a liquid droplet on the surface of a strongly wrinkled multilayer graphene sample, connected to two metallic electrodes. When the droplet is set in horizontal motion (at a velocity $\v_{\rm drop}$), an electric current is generated in the sample (see Fig. 6a). The roughness-induced friction force (per unit area) can be estimated as $ \mathbf{F}_{\rm drop} =  \lambda_0 \v_{\rm drop}$, 
with $\lambda_{\rm drop} \approx 3\cdot 10^6$ N.s.m${}^{-3} \times W$ \cite{Marcotte2022}, where $W \approx 10$ is a dimensionless parameter (dubbed wrinkling number) accounting for the wrinkle density. We note that $\v_{\rm drop}$ may be different from the interfacial velocity $\v_{\ell}$ used as an input parameter in our theory, depending on the hydrodynamic flow profile within the drop. However, we do not need to explicitly determine $\v_{\ell}$ since the wind velocity $\v_{\rm e}$ is dominated by the phonon wind $\v_{\rm ph}$. Because of the strong wrinkling, we choose to model the multilayer graphene sample as a zero-gap semiconductor with effective mass $m^* = 0.1 m_e$, $m_e$ being the electron mass. Then, since the experimental electronic density remains low, $\v_{\rm e} \approx \v_{\rm ph} /2 \approx \tb{F}_{\rm drop} / (2 \lambda_{\rm um})$. At a chemical potential $\mu = 20$ meV, the experimentally measured electro-fluidic conductivities are in the range 2-8 nA.s.$\mu$m${}^{-2}$, which is reproduced by our theoretical prediction (eq.~\eqref{currentgeneral}), for the wrinkling number $W$ in the range 5 - 20 (see Fig. 6b). Experimentally, the chemical potential is set by the bias $\Delta V$ applied between a grounded electrode and a working electrode: $\mu \approx \Delta V / 2$. Thus, the zero-gap semi-conductor model accounts for the experimentally observed linear scaling of $\sigma_{\rm ef}$ with the DC bias voltage.}

Overall, our theory is quantitatively consistent with the experiments of ref.~\cite{Marcotte2022} and is able to account for the particularly strong Coulomb drag currents (in the 10 nA range) that were generated by the motion of a micrometer-sized droplet at a few $\mu \rm m/s$. The key factor that is responsible for a strong current is the hydrodynamic friction force, which is large {in the experiments of ref.~\cite{Marcotte2022}} due to the high viscosity of the liquids used and the wrinkling of the sample surface: {this results in an efficient transfer of} a significant amount of momentum to the sample's phonon modes. The crucial role of phonons is further supported by the fact that the effect could be observed with a non-ionic silicon oil, where liquid-electron Coulomb interactions are negligible.  

\section{Quantum feedback and current-induced negative friction}

\begin{figure*}
\centering \includegraphics[scale=1]{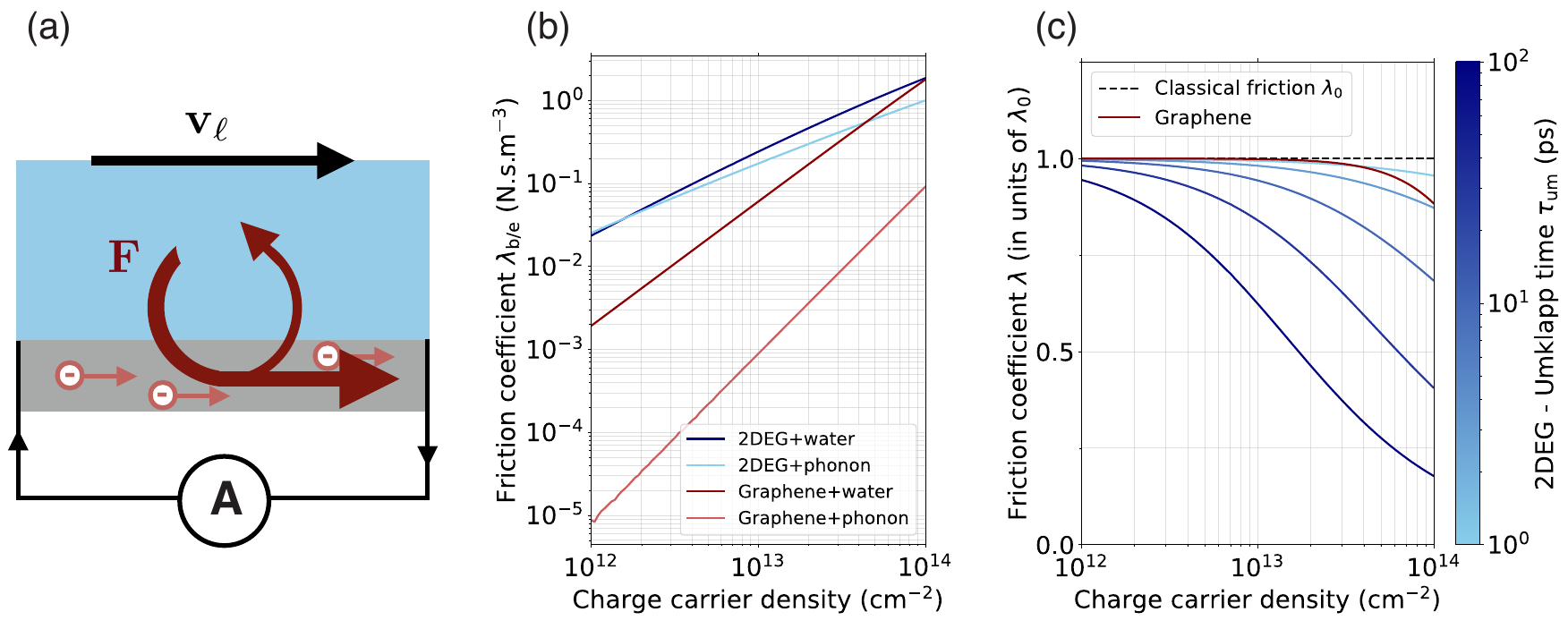}
\caption{\textbf{Quantum feedback and \rev{current-induced} negative friction.} (a) Schematic representation of the quantum feedback phenomenon: a fraction of the momentum transferred to the solid by classical friction is returned to the fluid by the solid's internal excitations. (b) Electron-boson friction coefficients as a function of the electron density for a 2DEG with $m = 0.1~\rm  m_{\rm e}$ and graphene, evaluated according to~\eqref{lambda_boson}. The calculations has been carried out using the electronic susceptibilities in the zero-temperature limit and in the RPA approximation.  (c) Hydrodynamic friction coefficient $\lambda$ (in units of the classical contribution $\lambda_0$) for a 2DEG with $m=0.1 m_{\rm e}$ and graphene.}
\label{fig7}
\end{figure*}

\subsection{Derivation}

Our analysis in Sec.~IV revealed that the hydrodynamic Coulomb drag current is determined by a subtle combination of electron-phonon and electron-hydron interactions. Indeed, the electron-hydron scattering provides a supplementary momentum relaxation pathway for the electrons, which prevents them from aligning to the phonon wind velocity. This immediately implies that the electrons actually transfer momentum \emph{to} the flowing liquid: they make a \emph{negative} contribution to the hydrodynamic friction force, as shown schematically in Fig. \ref{fig7}. In this section, we explicitly evaluate this negative contribution and assess its practical consequences.  

The liquid interacts with the solid through the classical roughness-induced friction force $\tb{F}_0$, and through the phononic and electronic contributions to the fluctuation-induced (quantum) friction $\tb{F}_{\rm h/ph}$ and $\tb{F}_{\rm h/e}$, respectively. The total hydrodynamic friction coefficient $\lambda$ is then defined according to 
\beq
\lambda\mc{A} \v_{\rm \ell} = \tb{F}_{0}+\tb{F}_{\rm h/ph}+\tb{F}_{\rm h/e}. 
\eeq
Introducing the individual friction coefficients as in Sec. IV B (see Table 1 for notations),  
\beq
\lambda\v_{\rm \ell} = \lambda_{0} \v_{\ell} +\lambda_{\rm h/ph} (\v_{\ell} - \v_{\rm ph}) +\lambda_{\rm h/e}(\v_{\ell} - \v_e).
\eeq
In the limit where $\lambda_0$ dominates all other friction coefficients, and using eqs. \eqref{ve2} and \eqref{vph2} for the velocities $\v_{\rm e}$ and $\v_{\rm ph}$, a rearrangement yields: 
\beq 
\lambda = \lambda_0 - \delta \lambda\quad\tn{with}\quad\delta\lambda=\frac{\lambda_0}{1+\frac{\lambda_{\rm um}+\frac{\tau}{\tau_{\rm im}}\lambda_{\rm ph/e}}{\lambda_{\rm h/ph}+\frac{\tau}{\tau_{\rm ph/e}}\lambda_{\rm h/e}}}, 
\label{delta_lambda}
\eeq
where we have included the possibility for the electrons to lose momentum through impurity scattering (at a rate $\tau_{\rm im}^{-1}$). This is our second key result, anticipated in Sec. II. Strikingly, $\delta \lambda$ represents a negative contribution to hydrodynamic friction. It is, however, always smaller than $\lambda_0$, so that the total friction coefficient remains positive, and there is no violation of the laws of thermodynamics. But a key observation is that it becomes equal to $\lambda_0$ in the absence of impurity or umklapp scattering. This amounts to formally considering a solid that is unable to relax momentum: then, even if the solid has a rough surface, all the momentum the liquid loses through the classical friction $\tb{F}_0$ is sent back by the solid's electronic and phononic fluctuations, so that the total friction vanishes. In practice, however, there is always some amount of momentum relaxation that keeps the friction from vanishing. The \rev{net reduction in hydrodynamic friction} is ultimately obtained by balancing the relaxation processes (umklapp and impurity scattering), and the processes that allow the solid to return momentum to the liquid: the phonon-hydron and electron-hydron interactions, the latter corresponding to quantum friction~\cite{Kavokine2022}.

In order to go beyond a qualitative discussion, we evaluate the electron-hydron and electron-phonon friction coefficients in the framework of ref.~\cite{Kavokine2022}. A general electron-boson friction coefficient is given by 
\beq
\lambda_{\rm b/e} =\frac{\hbar^2}{8\pi^2 T}\int_0^\infty\frac{\dd\omega\dd q}{\tn{sinh}^2\left(\frac{\hbar\omega}{2T}\right)} \,q^3\,\frac{\im[\chi_{\rm e}^{\rm R}]\im[D^{\rm R}]}{|1-\chi_{\rm e}^{\rm R}D^{\rm R}|^2},
\label{lambda_boson}
\eeq
where $D^{\rm R}$ is the retarded bosonic propagator in which the electron-boson interaction has been absorbed (see Sec. III). 
We evaluate Eq.~\eqref{lambda_boson} by numerical integration for both the graphene and 2DEG model systems, using zero temperature RPA expressions for the electronic susceptibility~\cite{Wunsch2006,Mihaila2011}. The results are plotted in Fig. \ref{fig7}b as a function of the electronic density. The relative values of the friction coefficients are consistent with the respective scattering times: indeed, one roughly expects $\lambda_{\rm b/e} \propto \tau^{-1}_{\rm b/e}$ (see Appendix C3). In particular, in the graphene case $\lambda_{\rm ph/e} \ll \lambda_{\rm um} \approx 2 ~\rm N \cdot s \cdot m^{-3}$ for all reasonable electronic densities, while in the 2DEG model this is true only for low enough electronic densities. 

In any case the term proportional to $\lambda_{\rm ph/e}$ in Eq.~\eqref{delta_lambda} is negligible in the limit of low impurity scattering: $1/\tau_{\rm im} \to 0$. If we further assume $\lambda_{\rm h/ph} \ll \frac{\tau}{\tau_{\rm ph/e}}\lambda_{\rm h/e}$ following the argument of Sec. IV B, Eq.~\eqref{delta_lambda} simplifies to 
\beq 
\lambda = \lambda_0  - \delta \lambda\quad\tn{with}\quad\delta\lambda=\frac{\lambda_0}{1+  \frac{\tau_{\rm ph/e}}{\tau} \frac{\lambda_{\rm um}}{\lambda_{\rm h/e}}}. 
\label{delta_lambda2}
\eeq
The resulting estimate for $\delta \lambda$ is plotted in Fig. \ref{fig7}c. In doped graphene, with an electronic density $10^{14}~\rm cm^{-2}$, we find that the negative quantum contribution reduces the hydrodynamic friction by about 12\%. In a 2DEG with parabolic dispersion, the quantum contribution to friction is stronger, so that at the same electronic density (and taking the graphene value for $\lambda_{\rm um}$) the friction reduction is nearly 80\%. 

\subsection{Discussion}

According to Eq.~\eqref{delta_lambda2}, the negative contribution to friction represents a \emph{quantum feedback} of the solid on the liquid flow: it originates from electronic excitations returning momentum that the liquid has lost due to the surface roughness. It is significant if the quantum friction coefficient $\lambda_{\rm h/e}$ is non-negligible compared to the umklapp friction coefficient. This is not a very stringent requirement: it is in fact satisfied for our two model systems, at least in part of the considered electron density range. 

The negative quantum friction effect is thus predicted to be quantitatively important. 
It complements the picture of quantum hydrodynamic friction developed in ref.~\cite{Kavokine2022}, where the solid was assumed to remain in an equilibrium state. This was justified in particular for a truly semi-infinite solid, where the surface can quickly transmit momentum to the bulk. In a 2D material that is weakly coupled to the underlying substrate, we find that momentum accumulation and the resulting Coulomb drag current are non-negligible. As electrons receive momentum from the liquid through the phonon-mediated interaction, they begin to move faster than the liquid, thus helping it flow along the surface and reducing the total hydrodynamic friction force. In this way, quantum friction can play a significant role even on non-atomically-flat surfaces. 

Negative quantum friction may provide a clue as to why the experimentally measured water friction on graphene~\cite{Xie2018} appears to be much lower than what is predicted by essentially all molecular simulations~\cite{Kannam2013,Tocci2014,Tocci2020,Thiemann2022}. Even in ab initio simulations at the Born-Oppenheimer level, it is not possible to account for Coulomb drag or fluctuation-induced quantum friction. Nevertheless, our quantitative estimates cannot account for the full extent of the discrepancy, and the specific case of water on pristine graphene will be the subject of further investigation. 

Ultimately, negative quantum friction provides a previously unexplored pathway for designing surfaces with low hydrodynamic friction. Friction reduction is first achieved by minimizing momentum relaxation in the solid: the best case scenario is a suspended 2D material, or a 2D material weakly coupled to its substrate. Then, the water-solid quantum friction coefficient needs to be maximized, so as to allow for efficient momentum transfer back to the liquid. Conditions for high quantum friction have been detailed in ref.~\cite{Kavokine2022}; typically, quantum friction benefits from high electronic densities and large effective masses. As outlined above, for the momentum transfer to be efficient, the quantum friction coefficient needs to be large compared to $\lambda_{\rm um}$, which is several orders of magnitude smaller than a typical hydrodynamic friction coefficient: many materials may satisfy this requirement. Potential candidates include magic angle twisted bilayer graphene~\cite{Stauber2016}, and metallic transition metal dichalcogenides such as $\rm VS_2$ and $\rm TaS_2$~\cite{DaJornada2020}.

\section{Conclusion: the importance of Quantum Interfacial Effects}

In this work, we introduce a novel perspective on hydrodynamics at solid-liquid interfaces. The usual description of such interfaces relies on the continuum Navier-Stokes equation, or -- going down to the molecular scale -- on the interactions between fluid molecules and the surface corrugation. This approach becomes insufficient to account for more advanced quantum couplings that arise between the liquid and the solid. 

Specifically, we start here by addressing the question of electronic current generation by interfacial liquid flows. Using the non-equilibrium Keldysh framework, we systematically investigate the mechanisms that couple fluid motion to the electronic degrees of freedom within the solid material, which include Coulomb and phonon-mediated interactions. The theory provides quantitative estimates for the electro-fluidic conductivity -- defined as the electronic current response to the fluid motion --, which are fully corroborated by experimental reports. 

Going further, our theoretical framework reveals a quantum feedback mechanism, that provides a negative contribution to hydrodynamic friction: in a very counter-intuitive way, quantum effects may reduce hydrodynamic friction at the solid-liquid interface. The Coulomb drag current can in fact be "faster" than the liquid flow, and return momentum to the liquid through electron-hydron scattering. The resulting \emph{negative quantum friction} provides a unique opportunity to tune hydrodynamic friction by choosing specific electronic properties of the confining wall. This broadens the scope of solid-liquid quantum friction beyond the water-carbon interfaces discussed in ref.~\cite{Kavokine2022}, as we find that it can play a role even for materials with non-negligible surface roughness. 

More generally, our results provide a new way of thinking about the interaction of liquids with solids, and in particular water-solid interfaces. By bridging fluid dynamics and condensed matter theory, we picture the interface dynamics in terms of the collective excitations of both the liquid and the solid, instead of real-space molecular interactions. Water charge fluctuations -- which we dub \emph{hydrons} -- couple to the quantum excitations inside the confining solid as first proposed in ref.~\cite{Kavokine2022}. Here, this approach bears fruit by accounting for existing experimental results, and predicting a novel quantum feedback effect. It opens the way to quantum engineering of fluid transport: quantum effects can become valuable assets for future water technologies.

The emerging interface between hydrodynamics, electrodynamics, condensed matter physics and quantum mechanics, is an uncharted territory that begs for further exploration.

\section*{Acknowledgements}

The authors acknowledge fruitful discussions with A. Marcotte, M. Lizee and A. Siria. The Flatiron Institute is a division of the Simons Foundation.  B.C. acknowledges funding from a J.-P. Aguilar grant of the CFM Foundation, and thanks Olivier Parcollet for hosting him at the Flatiron Institute. \rev{N.K. acknowledges support from a Humboldt fellowship}. L.B. acknowledges funding from the EU H2020 Framework Programme/ERC Advanced Grant agreement number 785911-Shadoks. 

\newpage 

\appendix 

\section{Useful results in the Keldysh formalism}

A reader interested in an extensive description of the formalism may consult \cite{Rammer2007, Kamenev2011, DiVentra2008}.

\subsection{Fluctuation-dissipation theorem}\label{appendix_FDT}

We consider particles (bosons or fermions) with creation and annihilation operators $\psi^\dagger(\tb{r},t)$ and $\psi(\tb{r},t)$, respectively. We define the real-time Green's functions
\beq
\left\{ \begin{array}{l}
G^{<}(\tb{r},t,\tb{r}',t')=\mp i\langle\psi^\dagger(\tb{r}',t')\psi(\tb{r},t)\rangle\\
G^{>}(\tb{r},t,\tb{r}',t')=-i\langle\psi(\tb{r},t)\psi^\dagger(\tb{r}',t')\rangle
\end{array}\right.
\eeq
where the upper sign is for the bosons. At equilibrium, the mean value of an operator $A$ is defined according to 
\beq \label{average}
\langle A\rangle=\frac{1}{\tr\left(e^{-\frac{\Ha}{T}}\right)}\tr\left(e^{-\frac{\Ha}{T}}A\right),\eeq
where $\Ha$ is the total Hamiltonian. 

Using the cyclicity of the the trace we deduce that at equilibrium
\beq 
G^{<}\left(\tb{r},t+\frac{i\hbar}{T},\tb{r}',t'\right)=\pm G^{>}(\tb{r},t,\tb{r}',t').\eeq
Upon Fourier transformation,
\beq \label{FD0}
G^{<} (\tb{q},\omega)=\pm e^{-\hbar \omega/T}G^{>}(\tb{q},\omega).\eeq
In terms of the R, A, K components used throughout the main text, 
\beqa\label{Gpm}
\left\{ \begin{array}{l}
G^{\rm R}-G^{\rm A}= G^{>}-G^{<}\\
G^{\rm K}=  G^{>}+G^{<}
\end{array}\right.
\eeqa 
From Eqs. \eqref{FD0} and \eqref{Gpm} we deduce the fluctuation-dissipation theorem:
\beq 
G^{\rm K}(\tb{q},\omega)=2i \frac{1\pm e^{-\hbar\omega/T}}{1\mp e^{-\hbar\omega/T}}\im\left[G^{\rm R}(\tb{q},\omega)\right].\eeq
This proves Eqs. \eqref{GKeq} and \eqref{DFD} of the main text. 

\subsection{Density and current}\label{appendix_current}

The particle density is given by $n(\tb{r},t)=\langle\psi^\dagger(\tb{r},t)\psi(\tb{r},t)\rangle$, that is, $n(\tb{r},t)=\pm iG^{<}(\tb{r},t,\tb{r},t)$. Using Eq. \eqref{Gpm}, the density can be expressed as 
\beq n=\pm \frac{i}{2}\left(G^{\rm K}-G^{\rm R}+G^{\rm A}\right)
\eeq
which yields Eq. \eqref{ne}. Let us notice that in real space and at equal times $G^{\rm R}(\tb{r},t,\tb{r}',t)-G^{\rm A}(\tb{r},t,\tb{r}',t)$ reduces to a constant.

Let us now derive an expression for the average electric current, forgetting for simplicity the spin degree of freedom. In the Heisenberg picture, the one-particle Hamiltonian is of the form 
\beq \mc{H}(t)=\int\dd \tb{r}\, \psi^\dagger(\tb{r},t) H_\tb{r}\psi(\tb{r},t)\eeq
where $H_\tb{r}$ is a differential operator acting in real space. Going to momentum space, this operator becomes a function of the quasi-momentum $\q$ of the electron. Within a given band, 
\beq \mc{H}(t)=\int \dd\tb{q}\, \psi^\dagger_\tb{q}(t) \hbar\xi_\tb{q}\psi_\tb{q}(t).\eeq
The Hamiltonian determines the dynamics of the density operator: 
\beqa \partial_t n_{\rm e}(\tb{r},t)&=& \frac{1}{i\hbar }[\psi^\dagger(\tb{r},t)\psi(\tb{r},t),\mc{H}(t)]\\
&=&\frac{1}{2\hbar } \left(H_\tb{r}-H_{\tb{r}'}\right)G^{\rm K}(\tb{r},t,\tb{r}',t)|_{\tb{r}=\tb{r}'}.\eeqa

On the other hand the electronic density satisfies the conservation equation
\beq e\partial_t n_{\rm e}(\tb{r},t)+\nabla\cdot\langle\tb{j}(\tb{r},t)\rangle=0.\eeq
from which we deduce
\beqa \langle\tb{j}(\tb{r},t)\rangle=\frac{e}{2\hbar } \left([H_\tb{r},\tb{r}]-[H_{\tb{r}'},\tb{r}']\right)G^{\rm K}(\tb{r},t,\tb{r}',t)|_{\tb{r}=\tb{r}'}. ~~~~~~~~ \eeqa
Going to Fourier space and assuming translational invariance in time and space, we obtain
\beqa \langle\tb{j}\rangle=ie \int\frac{\dd\tb{q}\dd\omega}{(2\pi)^3}\,\left(\nabla_\q \xi_\q\right) \, G^{\rm K}(\tb{q},\omega).\eeqa
Multiplying by 2 to account for the spin degeneracy we recover Eq. \eqref{current0} of the main text. An additional factor can be included to account for a valley degeneracy.

\section{Evaluation of the self-energy}\label{appendix_selfenergy}

\subsection{General expression}\label{appendix_selfenergy1}

In this appendix we evaluate the electron-boson self-energy diagram in Fig.~\ref{fig2}b. We compute the self-energy in a given electronic band with dispersion $\xi_{\q}$ and eigenstates $|\q \rangle$, and we neglect interband electron scattering: this is reasonable as long as the boson energy is small compared to the Fermi level. 
Applying the Keldysh formalism Feynman rules in real space, we find \cite{Rammer2007}:
\beq\label{Sigma1}
\left\{
\begin{array}{l}
\Sigma^{\rm R,A} = \dfrac{i}{2} \left( D^{\rm R,A} G^{\rm K}_0 + D^{\rm K} G^{\rm R,A}_0\right) \\
\\
\Sigma^{\rm K} = \dfrac{i}{2} \left(G^{\rm R}_0-G^{\rm A}_0\right)\left(D^{\rm R}-D^{\rm A}\right) + \dfrac{i}{2} D^{\rm K} G^{\rm K}_0 
\end{array}
\right.  
\eeq
Starting with the retarded component, the products become convolutions in Fourier space: 
\begin{widetext}
\begin{equation}
\Sigma^{\rm R} (\q,\omega) = \frac{i}{2} \int \frac{\d \q' \d \omega'}{(2 \pi)^3} \mathcal{M}(\q-\q',\q) \left[D^{\rm R}(\q',\omega'-\q'\!\cdot\! \v_{\rm b}) G_0^{\rm K}(\q-\q',\omega-\omega') + D^{\rm K}(\q',\omega'-\q'\!\cdot\! \v_{\rm b}) G_0^{\rm R}(\q-\q',\omega-\omega') \right],
\end{equation}
where $\mathcal{M}(\q-\q',\q) \equiv |\langle \q - \q' | e^{-i\q' \r} | \q \rangle |^2$. Using Eqs.~\eqref{GReq} and \eqref{GKeq} for the equilibrium Green's functions, as well as the bosonic fluctuation-dissipation theorem in Eq.~\eqref{DFD}, we obtain 
\begin{equation}
\Sigma^{\rm R} (\q,\omega) = \frac{i}{2} \int \frac{\d \q' }{(2\pi)^3} \mathcal{M}(\q-\q',\q) \left[ \int \d \omega' 2i f(\omega'-\q'\!\cdot\! \v_{\rm b})\frac{ \mathrm{Im} \, [ D^{\rm R}(\q',\omega'-\q' \!\cdot\!\v_{\rm b})]}{\omega-\omega'- \xi_{\q-\q'} + i 0^+} - \frac{2 i \pi}{f(\xi_{\q-\q'})} D^{\rm R}(\q',\omega - \xi_{\q-\q'} - \q'\!\cdot\! \v_{\rm b}) \right]. 
\label{SigmaR1}
\end{equation}
Introducing the Lehmann representation for the bosonic propagator, 
\begin{equation}
D^{\rm R}(\q,\omega) = -\frac{1}{\pi} \int_{-\infty}^{+\infty} \d \omega'  \frac{\mathrm{Im} \, [D^{\rm R}(\q,\omega')]}{\omega - \omega' + i 0^+},
\end{equation}
Eq.~\eqref{SigmaR1} becomes 
\beq\label{SigmaR}
\Sigma^{\rm R}(\q,\omega) = -\frac{1}{\hbar} \int \frac{\d \q' \d \omega'}{(2\pi)^3} \mathcal{M}(\q-\q',\q) \frac{ \mathrm{Im} \, [ D^{\rm R}(\q',\omega'-\q'\!\cdot\!\v_{\rm b})]}{\omega-\omega'- \xi_{\q-\q'} + i 0^+} \left( f(\omega' -\q' \!\cdot\!\v_{\rm b}) + \frac{1}{f(\xi_{\q-\q'})} \right)
\eeq
which is Eq. \eqref{SR} of the main text. Following these exact same steps, we may check that $\Sigma^{\rm A}(\q,\omega) = \Sigma^{\rm R}(\q,\omega)^*$. Making use again of fluctuation-dissipation relations, we find for the Keldysh component of the self-energy 
\begin{equation}
\Sigma^{\rm K}(\q,\omega) = -2i \int \frac{\d \q' \d \omega'}{(2\pi)^3} \mathcal{M}(\q-\q',\q)\left( 1+ \frac{f(\omega'-\q'\!\cdot\! \v_{\rm b}) }{f(\omega-\omega')} \right) \Im{D^{\rm R}(\q',\omega'-\q'\v)} \Im{G^{\rm R}(\q-\q',\omega-\omega')}.
\end{equation}
Using that $\Im{G^{\rm R}(\q,\omega)} = - \pi \delta(\omega-\xi_{\q})$, we obtain 
\begin{equation}
\Sigma^{\rm K}(\q,\omega) = i \int \frac{\d \q' }{(2\pi)^2}\mathcal{M}(\q-\q',\q) \left( 1+ \frac{f(\omega-\xi_{\q-\q'} - \q'\!\cdot\! \v_{\rm b}) }{f(\xi_{\q-\q'})} \right) \Im{D^{\rm R}(\q',\omega-\xi_{\q-\q'}-\q'\!\cdot\! \v_{\rm b})},
\label{SigmaK2}
\end{equation}
which is Eq. \eqref{SK} of the main text.
\end{widetext}

\subsection{Bosonic propagators}

The expressions for the boson (hydron and phonon) propagators that are relevant for our model systems are given in the main text (Eqs.~\eqref{Dwater} - \eqref{Dphonon}). Here, we provide a few additional details, in particular concerning the derivation of the hydron propagator. Starting from the electron-hydron Hamiltonian in Eq.~\eqref{Heh}, the bare hydron propagator is defined as 
\begin{equation}
D^{\rm R}_{\rm w,0} (\q,t) =  -i  \theta(t)(V^{\rm C}_{\q})^2 \langle [ n_{\rm s} (\q,t), n_{\rm s}(-\q,0) ] \rangle ,
\end{equation}
with 
\beq
n_{\rm s}(\q,t) = \int \d \boldsymbol{\rho}  \int_0^{+\infty} \d z e^{-i \q \boldsymbol{\rho}} e^{-qz} n_{\rm \ell}(\r,t).
\eeq
We have isolated here the component of the position $\r$ that is perpendicular to the interface: $\r = \boldsymbol{\rho} + z \mathbf{e}_z$. Identifying the water density-density response function $\chi^{\rm R}_{\rm w}(\q,z,z',\omega)$, we obtain 
\begin{equation}
\begin{split}
D^{\rm R}_{\rm w,0} (\q,\omega) &= (V_{\q}^{\rm C})^2 \int_0^{+\infty} \d z \d z' e^{q (z+z')} \chi^{\rm R}_{\rm w}(\q,z,z',\omega) \\
&\equiv - V_{\q}^{\rm C} g^{\rm R}_{\rm w}(\q,\omega),
\end{split}
\end{equation}
recovering the water surface response function $g^{\rm R}_{\rm w}(\q,\omega)$ that was studied extensively in ref.~\cite{Kavokine2022}. It was found to be well described by the sum of two Debye peaks: 
\beq
g^{\rm R}_{\rm w}(\q,\omega) = \sum_{k = 1,2} \frac{f_k}{1- i  \omega/\omega_{\mathrm{D},k} },
\eeq
with $\omega_{\rm D,1} \sim 1~\rm meV$ and $\omega_{\rm D,2} \sim 100~\rm meV$, and $f_1 \approx f_2 \sim 0.5$. Accounting for the RPA screening as per the diagrams in Fig. 2c-d, the full hydron propagator becomes 
\beq
D_{\rm w}^{\rm R} (\q,\omega) = \frac{D^{\rm R}_{\rm w,0} (\q,\omega) (1 + V_{\q}^{\rm C} \chi^{\rm R}_{\rm e} (\q,\omega))^2}{1 - D^{\rm R}_{\rm w,0}(\q,\omega) \chi^{\rm R}_{\rm e}(\q,\omega)}, 
\eeq
where $\chi_{\rm e}^{\rm R}$ is the electronic density-density response function. Using the definition of the dielectric function $\epsilon(\q,\omega) = 1/(1-V_{\q}^{\rm C} \chi_{\rm e}^{\rm R} (\q,\omega))$, and neglecting its frequency dependence on the scale of the Debye frequencies, we recover Eq.~\eqref{Dwater} of the main text. 

For acoustic phonons in the framework of a jellium model, the fully screened propagator is \cite{Bruus2004}: 
\begin{equation}
D_{\rm ph}^{\rm R}(\q,\omega) = \frac{1}{\hbar}\frac{V_q^{\rm C}}{\epsilon(q)} \frac{\omega_{\q}^2}{(\omega+i0^+)^2-\omega_{\q}^2}, 
\end{equation}
with $\omega_{\q} = cq$, $c$ being the sound velocity. The electron-phonon interaction is essentially a screened Coulomb interaction. Conversely, in graphene, the electron-phonon interaction has a peculiar form~\cite{Ochoa2011}, so that 
\begin{equation}
D_{\rm ph}^{\rm R}(\q,\omega) = \frac{1}{\hbar}V^{\rm ph/e} \frac{\omega_{\q}^2}{(\omega+i0^+)^2-\omega_{\q}^2}, 
\end{equation}
with $V^{\rm ph/e}= g^2/(2 \rho c^2)$; here, $g\approx 3$ eV, is the electron-phonon coupling and $\rho\approx 7.6\e{-7}$ kg/m${}^2$ is the mass per unit area.

In the 2DEG, the matrix elements $\mathcal{M}$ are unity. We use the Thomas-Fermi approximation for the dielectric function: $\epsilon(q) = 1+ q_{\rm TF}/q. $
Since the density of states is independent of energy, the Thomas-Fermi wavevector $q_{\rm TF}$ does not depend on the Fermi level: $q_{\rm TF} = 2/a_0$, with $a_0 = 4\pi \epsilon_0 \hbar^2/(m e^2)$ the Bohr radius at the effective mass. 
In graphene, $\mathcal{M}(\q-\q',\q) = \frac{1}{2} (1 + \cos \theta_{\q-\q',\q})$, where $ \theta_{\q-\q',\q}$ is the angle between $\q-\q'$ and $\q$~\cite{Hwang2007}. For the RPA dielectric function, a full analytical expression can be found in the literature~\cite{Hwang2007,Wunsch2006}. 

\subsection{Impurity approximation}\label{appendix_selfenergy2}

The self-energies can be simplified within the impurity approximation, which amounts to taking the limit of vanishing bosonic frequency. This is expected to be reasonable as long as the bosonic frequencies are much lower than $T/\hbar$.
For the hydron propagator, as $\omega_{\rm D, 1, 2} \to 0$, 
\begin{equation}
f(\omega) \Im{D^{\rm R}_w(\q,\omega)} \to - 2\pi \frac{T}{\hbar^2}\frac{V_{\q}^C}{\epsilon(\q)} \delta(\omega).
\end{equation}
For the acoustic phonons, as $\omega_{\q} \to 0$,
\begin{equation}\label{Dph}
f(\omega) \Im{D^{\rm R}_{\rm ph}(\q,\omega)} \to -2\pi \frac{T}{\hbar^2}\frac{V_{\q}^{\rm ph/e}}{\epsilon(\q)} \delta(\omega).
\end{equation}
Thus, within the impurity approximation, the hydron and phonon propagators become formally identical. 

As the bosonic frequencies are taken to 0, the terms that are not proportional to $f(\omega)$ in Eq. \eqref{SigmaR} become negligible. We then obtain 
\begin{equation} 
\Sigma^{\rm R}(\q,\omega) = \frac{T}{\hbar^2} \int \frac{\d \q'}{(2\pi)^2}  \frac{\mathcal{M}(\q',\q) V_{\q-\q'}}{\omega-(\q-\q') \!\cdot\!\v_{\rm b} - \xi_{\q'} + i 0^+}, 
\end{equation}
with $V = V^{\rm C}_{\q}/\epsilon(q)$ or $V^{\rm ph/e}$. In the usual treatment of impurity scattering~\cite{Bruus2004}, one further neglects the real part of the self-energy, and the frequency dependence of the imaginary part: 
$\mathrm{Im} \, [\Sigma^{\rm R} (\q,\omega) ] \approx - 1/ \tau_{\q} \equiv \mathrm{Im} \, [\Sigma^{\rm R} (\q,\xi_{\q}) ]$. The quasiparticle scattering rate at wavevector $\q$ is given by 
\begin{equation}\label{selfenergy2appendix}
 \frac{1}{\tau_{\q}} =  \pi \frac{T}{\hbar^2} \int \frac{\d \q'}{(2\pi)^2} \mathcal{M}(\q,\q') V_{\q-\q'} \delta(\xi_{\q} - \xi_{\q'}) 
\end{equation}
which is Eq. \eqref{selfenergy2} of the main text. An explicit estimate can be obtained under the assumption $V_{\q-\q'} \approx V_{\q}$ (and $\mathcal{M}(\q,\q') = 1$). Then, by changing variables $\dd\q'=(2\pi)^2N(u)\dd u$, where $N(u)$ is the density of states at energy $u$, we deduce 
\begin{equation}\label{SEapproxappendix}
 \frac{1}{\tau_{\q}} \approx  \pi \frac{T}{\hbar}  V_{\q}N(u_\q)
\end{equation}
which is Eq. \eqref{SEapprox} of the main text.

\subsection{Fermi's golden rule}\label{appendix_FGR}

Eq.~\eqref{SEapproxappendix} can be obtained by writing down a simplified Fermi golden rule for an electron-phonon interaction of the form
\beq\label{Hephappendix}
\mc{H}_{\rm ph/e}(t) =g\mc{A}^2 \int\frac{\dd\tb{q}\dd\tb{k}}{(2\pi)^4}\, \psi^\dag_\q (t)b^\dag_\tb{k}(t)\psi_{\q+\tb{k}}(t)+\tn{h.c.} 
\eeq
where $g$ is the coupling constant and $b^\dag$ the creation operator of the phonon (h.c. stands for hermitian conjugate). Fermi's golden rule predicts:
\beq 1/\tau_\q\approx \frac{2\pi}{\hbar} g^2\mc{A}N(u_\q)  n_{\rm B}(\omega_{\q}). \eeq
Here, the Bose distribution $n_{\rm B}$ counts the number of modes on which the electrons can scatter. Assuming $\hbar\omega_{\q}\ll T$,
\beq 1/\tau_\q\approx \frac{2\pi}{\hbar} g^2\mc{A}N(u_\q) T/ \hbar\omega_{\q}. \eeq

With our usual notations the coupling constant $g$ is absorbed inside the effective potential $V_\q$, according to $V_\q=g^2\mc{A}/\epsilon(q)\omega_{\q}$. Therefore, we recover the estimate provided in Eqs. \eqref{SEapprox} and \eqref{SEapproxappendix} from the Fermi's golden rule.

For the electron-hydron interaction the computation is more involved since we would in principle have to consider the superposition of many hydron modes. Qualitatively, reproducing the above reasoning for each of the modes yields exactly the same result. 

\subsection{Higher order diagrams: extension of the Migdal theorem} \label{appendix_higherorder}

\begin{figure*}
\centering \includegraphics[scale=1]{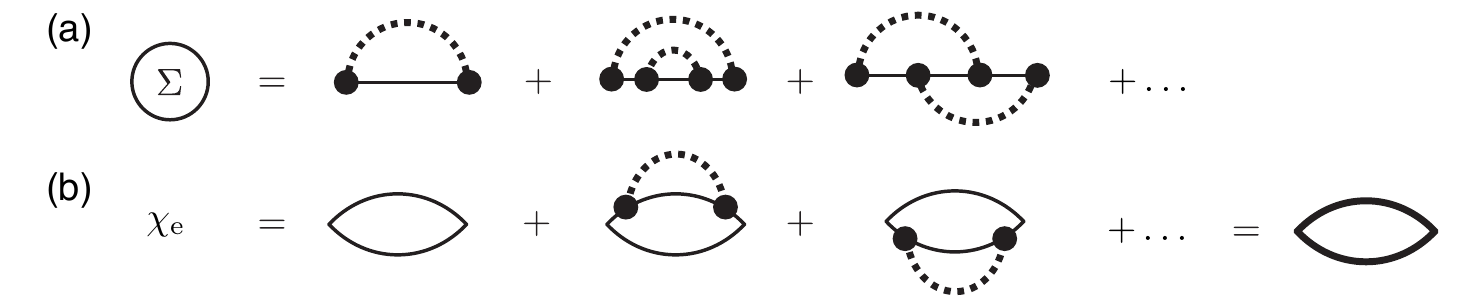}
\caption{\textbf{Diagrammatic expansions.} (a) First and second order diagrams for the electron-boson self-energy. (b) Diagrammatic representation of the electronic susceptibility in terms of the renormalized Green's functions (thick lines).}\label{fig8}
\end{figure*}

Up to now, we have only considered the first order diagram for the self-energy. Here, we provide an estimate for higher order diagrams and show that they can reasonably be neglected. The $n$-th order self-energy diagram contain $n$ bosonic propagators and $2n-1$ electronic propagators. Such a diagram includes $n$ loops, therefore there are $n$ momentum integrals and $n$ frequency integrals. The first few self-energy diagrams are displayed in Fig. \ref{fig8}a.

In the impurity approximation, the bosonic propagators take the form 
\beqa
\left\{ \begin{array}{l}
\im\left[D^{\rm R}(\q,\omega)\right]\approx -\pi\dfrac{\omega}{\hbar} V_\q \delta(\omega)=0\\
\\
D^{\rm K}(\q,\omega)\approx-4i\pi\dfrac{T}{\hbar^2} V_\q \delta(\omega)\end{array}\right.
\eeqa 
Therefore, in the expression of the self energy the only non-vanishing terms are those that only include the Keldysh component of the bosonic propagator. In the Keldysh diagrammatic rules each vertex must be summed over the dynamic indices. However, when restricting to the Keldysh propagator for the bosons, the dynamic index of the electronic Green function is preserved at each vertex \cite{Rammer2007}. We shall focus on the retarded component of the self energy. When imposing these external dynamic indices, the electronic propagators can only be retarded or advanced Green functions thanks to the trigonal structure. To obtain an estimate of the self energy we will neglect the real part of the retarded/advanced electronic Green function, which then reduces to its imaginary part: 
\beq
G^{\rm R,A}(\q,\omega)\approx \mp i\pi\delta(\omega-\xi_\q).
\eeq

Putting everything together, we have $n$ integrations over internal 2D momenta, $n$ integrations over internal frequencies, $n$ bosonic Keldysh Green functions, each of them introducing a Dirac distribution that fixes a frequency, and $2n-1$ electronic propagators, each of them introducing a Dirac distribution that fixes a momentum, and a factor $1/(\nabla_q\xi_\q)$. After integration of all the Dirac distributions there only remains an angular integration. The structure of the diagram determines the numerical factor coming from the angular integration, and the momentum at which the $n$ interactions $V_\q$ and the $2n-1$ velocities $\nabla_q\xi_\q$ are evaluated. To provide an order of magnitude, we evaluate them at the Fermi wavevector and take the angular factor equal to 1.
We then obtain the following order of magnitude for the $n$-th order self-energy diagram:
\beq
\Sigma^{(n)}\sim \left(\frac{TV_{k_{\rm F}}}{\hbar^2}\right)^n\frac{k_{\rm F}}{v_{\rm F}^{2n-1}}.
\eeq
For $n=1$ we recover the estimate for the first order self energy calculated above (Eq. \eqref{SEapproxappendix}).
Comparing the $n$-th and first order self-energies, 
\beq
\frac{\Sigma^{(n)}}{\Sigma^{(1)}}\sim \left(\frac{TV_{k_{\rm F}}}{\hbar^2v_{\rm F}^2}\right)^{n-1}\equiv \alpha^{n-1}.
\eeq
Let us estimate the coefficient $\alpha$ for the different models.

For the 2DEG, $V_{k_{\rm F}}$ is the screened the Coulomb potential. For reasonable electronic densities ($k_{\rm F}\ll 1/a_0$), the dielectric constant is $\epsilon(k_{\rm F})\approx \frac{2}{a_0k_{\rm F}}$ where $a_0$ is the Bohr radius computed at the effective mass. Thus, $V_{k_{\rm F}}\approx \hbar^2/m$ and $\alpha \sim T/\mu$. Therefore, neglecting higher order diagrams is valid as long as the chemical potential is large compared to the temperature. 

For graphene, $\epsilon(k_{\rm F})\approx \frac{e^2}{\epsilon_0 \hbar v_{\rm F}} \sim 1$. For the electron-hydron interaction, we obtain again $\alpha\sim T/\mu$. For the electron-phonon interaction, we find $\alpha\sim \frac{TV^{\rm eff}}{\hbar v_{\rm F}^2}\sim 10^{-4}$: higher order self-energy diagrams can always be neglected.

\section{Fluctuation-induced friction forces}

\subsection{Susceptibility renormalisation}\label{appendix_susceptibility}

In this section, we compute the non-equilibrium electronic susceptibility (density-density response function) $\boldsymbol{\chi}_{\rm e}$ starting from the electronic Green, according to the diagrammatic definition given in Fig. \ref{fig8}b. We neglect in particular vertex corrections due to the electron-boson interaction. This diagram is formally equivalent to the diagram for the first order self energy \cite{Kamenev2011}. Thus, the calculation is similar, if one replaces the bosonic propagator $\tb{D}$ by the \emph{backward} electronic Green's function $G_{\rm B}^{ij}(\tb{r}_1,t_1,\tb{r}_2,t_2)=G^{ji}(\tb{r}_2,t_2,\tb{r}_1,t_1)$, where $i,j$ are the Keldysh dynamical indices. For convenience we denote in the following $\check{F}(x)=F(-x)$. Using that in real space $\check{G}^{\rm R,A}=(G^{\rm A,R})^*$ and $\check{G}^{K}=-(G^{K})^*$ by construction, the backward propagators become in the Keldysh trigonal representation and in Fourier space:
\beqa\label{GB}
\left\{ \begin{array}{l}
G_{\rm B}^{\rm R,A}(\q,\omega)=(G^{\rm A,R})^*(-\q,-\omega)\\
G_{\rm B}^{K}(\q,\omega)=-(G^{K})^*(-\q,-\omega)
\end{array}\right.
\eeqa 
Let us notice that the backward electronic Green function satisfies the same fluctuation-dissipation theorem as the forward electronic Green's function $\tb{G}$. Using Eq. \eqref{Sigma1} and the fluctuation-dissipation theorem in Eq. \eqref{GFD}, we obtain \beq\label{chi1}
\left\{
\begin{array}{l}
\chi^{R}_{\rm e} = - \left( G_{\rm B}^{R} \ast \frac{\im[G^{\rm R}]}{f} + \frac{\im[G^{\rm R}_{\rm B}]}{f} \ast G^{R}\right) \\
\\
\chi^{\rm K}_{\rm e} = -2i\left( \im[G^{\rm R}_{\rm B}]\ast \im[G^{\rm R}] +  \frac{\im[G^{\rm R}_{\rm B}]}{f}\ast \frac{\im[G^{\rm R}]}{f} \right)
\end{array}
\right.  
\eeq
where at frequency $\omega$ and momentum $\q$, $f$ stands for $f(\omega-\q\cdot\v_{\rm e}(q))$. As long as $\q\cdot\v_{\rm e}(q) \ll \omega$, we may wompare these formulas point by point, and using the trigonometric identities 
$f^{-1}_{\omega'}+f^{-1}_{\omega-\omega'}=f^{-1}_{\omega}\frac{1-f^{-1}_{\omega'}}{1-f^{-1}_{\omega}f^{-1}_{\omega'}}$
 and $1+f^{-1}_{\omega'}f^{-1}_{\omega-\omega'}= \frac{1-f^{-1}_{\omega'}}{1-f^{-1}_{\omega}f^{-1}_{\omega'}}$,
  we deduce a the quasi-equilibrium fluctuation-dissipation theorem for the susceptibility:
\beq \label{chiFDappendix}
\chi_{\rm e}^{\rm K}(\tb{q},\omega)=2if(\omega-\tb{q}\!\cdot\!\tb{v}_{\rm e}(q))\im[\chi_{\rm e}^{\rm R}(\tb{q},\omega)]
\eeq
which is Eq. \eqref{chiFD} of the main text.

\subsection{Electron-boson friction force}\label{sec_force}

We now generalize the result of ref.~\cite{Kavokine2022} for fluctuation-induced quantum friction to account for the non-equilibrium state of the solid. 
The electron-boson force per unit surface can be expressed as~\cite{Kavokine2022}:
\beq\label{force}
\frac{\langle\tb{F}_{\rm b/e}\rangle}{\mc{A} } =\frac{\hbar}{4\pi}\int\frac{\dd\omega\dd\tb{q}}{(2\pi)^3} \,\tb{q}\,\chi_{\rm b/e}^{\rm K}(\tb{q},\omega).
\eeq
where the Keldysh cross correlation $\chi_{\rm b/e}^{\rm K}$ is given by
\beq \label{chieb1}
\chi^{\rm K}_{\rm eb}=-\frac{\chi_{\rm e}^{\rm A}D^{\rm K}+\chi_{\rm e}^{\rm K}D^{\rm R}}{|1-\chi_{\rm e}^{\rm R}D^{\rm R}|^2},
\eeq
with all the correlation function being computed in the non-equilibrium state. We now use the \emph{quasi-equilibrium} fluctuation-dissipation theorems in Eqs. \eqref{DFD} and Eq. \eqref{chiFDappendix}, to obtain 
\beq \label{chieb2}
\chi^{\rm K}_{\rm eb}=-2i\Delta f\frac{\im[\chi_{\rm e}^{\rm R}]\im[D^{\rm R}]}{|1-\chi_{\rm e}^{\rm R}D^{\rm R}|^2}.
\eeq
where $\Delta f=f(\omega-\tb{q} \cdot  \v_{\rm e})-f(\omega-\tb{q}\cdot\tb{v}_{\rm b})$. Therefore, we find that as long as the quasi-equilibrium condition holds, the electron-boson friction coefficient is computed as if the electrons were at equilibrium:
\beq\label{lambda_boson_appendix}
\lambda_{\rm b/e} =\frac{\hbar^2}{8\pi^2 T}\int_0^\infty\frac{\dd\omega\dd q}{\tn{sinh}^2\left(\frac{\hbar\omega}{2T}\right)} \,q^3\,\frac{\im[\chi_{\rm e}^{\rm R}]\im[D^{\rm R}]}{|1-\chi_{\rm e}^{\rm R}D^{\rm R}|^2}
\eeq 
which is Eq. \eqref{lambda_boson} of the main text. However, the friction force is now proportional to the differential velocity: $\langle\tb{F}_{\rm b/e}\rangle /\mc{A} = \lambda_{\rm b/e} (\v_{\rm b}-\tb{v}_{\rm e})$. 

\subsection{Scaling of the electron-boson friction coefficient}\label{sec_lambda}

In this section, we provide a qualitative approach to electron-boson friction, that is able to predict the scaling of the friction coefficient with electronic density and effective mass. In the reference frame where the bosons do not move, the electrons are subject to a wind velocity $\v_{\rm e}-\v_{\rm b}$. They relax by scattering on the bosons at a rate $\tau_{\rm b/e}^{-1}$. The force (or momentum flux) per unit surface is then given by
\beq -\frac{\tb{F}_{\rm b/e}}{\mc{A}}\sim \frac{\hbar k_{\rm F}\times N(\mu)\hbar k_{\rm F}(\v_{\rm e}-\v_{\rm b})}{\tau_{\rm b/e}}.\eeq
This is the momentum of an electron (at the Fermi level), times the number of electrons that are able to scatter (in a zero temperature picture), times the scattering rate. The scaling of the friction force with the inverse of the electron-boson scattering time is consistent with the relation between resistivity and electronic friction coefficient proposed by Persson~\cite{Persson1991,Persson_book}.

Therefore, using Eq. \eqref{SEapproxappendix}, we find that the electron-boson friction coefficient scales as
\beq
\lambda_{\rm b/e} \sim \pi \hbar k_{\rm F}^2 N(\mu)^2 TV_{k_{\rm F}}.
\eeq
Let us notice that in this approximation the electron-boson friction coefficient does not depend on the dynamics of the bosons but only on the electronic structure and the interaction potential: this is in fact the analogue of the impurity approximation for the friction coefficient. For a 2DEG with reasonable electronic density, the screened Coulomb potential is roughly independent of the Fermi level and the effective mass. Since the density of states $N(\mu)=m/\hbar^2$ is constant and $k_{\rm F}\sim\sqrt{m\mu}$ we expect $\lambda_{\rm b/e}\propto m^2\mu\propto mn$ where $n$ is the charge carriers density, for both the electron-phonon and electron-hydron interactions.  In graphene, the screened Coulomb potential scales as $1/k_{\rm F}\propto 1/\mu$, the density of states $N(\mu)=2\mu/(\pi\hbar^2 v_{\rm F}^2)$ scales as $\mu$ and $k_{\rm F}=\mu/v_{\rm F}$. Therefore, we expect $\lambda_{\rm b/e}\propto \mu^3\propto n^{3/2}$ for the electron-hydron interaction. On the other hand, using that the effective potential $V_\q$ does not depend on $q$ for the electron-phonon interaction in graphene, we expect $\lambda_{\rm b/e}\propto \mu^4\propto n^{2}$. We thus recover the scalings of the full numerical results displayed in Fig. \ref{fig7}b.



%

\end{document}